\documentclass[%
reprint,
superscriptaddress,
 amsmath,amssymb,
prl,
]{revtex4-2}

\usepackage{graphicx}
\usepackage{dcolumn}
\usepackage{bm}
\usepackage{hyperref}

\begin{document}


\preprint{APS/123-QED}

\title{Spectral-isolated photonic topological corner mode with a tunable mode area and stable frequency}

\author{Zhongfu Li}\thanks{These authors contributed equally to this work.}
 \affiliation{School of Science and Engineering, The Chinese University of Hong Kong, Shenzhen, Guangdong, 100190, China.}
 \affiliation{New Cornerstone Science Foundation, Department of Physics, The University of Hong Kong, Pokfulam Road, Hong Kong, 999077, China.}
 \author{Shiqi Li}\thanks{These authors contributed equally to this work.}
 \affiliation{Collaborative Innovation Center of Advanced Microstructures and School of Physics, Nanjing University, Nanjing, 210093, China.}
 
\author{Bei Yan}\thanks{These authors contributed equally to this work.}
\affiliation{Department of Electronic and Electrical Engineering, Southern University of Science and Technology, Shenzhen, Guangdong, China.}%

\author{Hsun-Chi Chan}\thanks{These authors contributed equally to this work.}
\affiliation{New Cornerstone Science Foundation, Department of Physics, The University of Hong Kong, Pokfulam Road, Hong Kong, 999077, China.}

\author{Jing Li}
\affiliation{School of Science and Engineering, The Chinese University of Hong Kong, Shenzhen, Guangdong, 100190, China.}

\author{Jun Guan}
\affiliation{School of Science and Engineering, The Chinese University of Hong Kong, Shenzhen, Guangdong, 100190, China.}

\author{Wengang Bi}
\affiliation{School of Science and Engineering, The Chinese University of Hong Kong, Shenzhen, Guangdong, 100190, China.}

\author{Yuanjiang Xiang}
\affiliation{School of Physics and Electronics, Hunan University, Changsha, 410082, China}

\author{Zhen Gao}%
 \email{gaoz@sustech.edu.cn}
\affiliation{Department of Electronic and Electrical Engineering, Southern University of Science and Technology, Shenzhen, Guangdong, China.}
\affiliation{State Key Laboratory of Optical Fiber and Cable Manufacture Technology, Southern University of Science and Technology, Shenzhen, Guangdong, China.}

\author{Shuang Zhang}
 \email{shuzhang@hku.hk}
\affiliation{New Cornerstone Science Foundation, Department of Physics, The University of Hong Kong, Pokfulam Road, Hong Kong, 999077, China.}

\author{Peng Zhan}
 \email{zhanpeng@nju.edu.cn}
\affiliation{Collaborative Innovation Center of Advanced Microstructures and School of Physics, Nanjing University, Nanjing, 210093, China.}

\author{Zhelin Wang}
\affiliation{Collaborative Innovation Center of Advanced Microstructures and School of Physics, Nanjing University, Nanjing, 210093, China.}

\author{Biye Xie}
 \email{xiebiye@cuhk.edu.cn}
\affiliation{School of Science and Engineering, The Chinese University of Hong Kong, Shenzhen, Guangdong, 100190, China.}



\begin{abstract}
Emergent collective modes in lattices give birth to many intriguing physical phenomena in condensed matter physics. Among these collective modes, large-area modes typically feature small-level spacings, while a mode with stable frequency tends to be spatially tightly confined. Here, we theoretically propose and experimentally demonstrate a spectral-isolated photonic topological corner mode with a tunable mode area and stable frequency in a two-dimensional photonic crystal. This mode emerges from hybridizing the large-area homogeneous mode and in-gap topological corner modes. Remarkably, this large-area homogeneous mode possesses unique chirality and has a tunable mode area under the change of the mass term of the inner topological non-trivial lattice. We experimentally observe such topological large-area corner modes(TLCM) in a 2D photonic system and demonstrate the robustness by introducing disorders in the structure. Our findings have propelled the forefront of higher-order topology research, transitioning it from single-lattice systems to multi-lattice systems. They may support promising potential applications, particularly in vertical-cavity surface-emitting lasers.
\end{abstract}

\maketitle

{\it{Introduction}}.---{Attaining large-area\cite{chenHigh2017,liuPeriodic2016} and robust\cite{eatonLasing2016} semiconductor lasers is one of the most pivotal goals in laser physics. Traditionally, as photonic modes occupy finite-size devices, expanding the mode area tends to reduce modes' level spacings\cite{eatonSemiconductor2016,gaoDiracvortex2020,schawlowInfrared1958,kogelnikSTIMULATED1971,sodaGaInAsP1979,imadaCoherent1999,yoshidaDoublelattice2019,li2024CZM}. Recently, Chua and Lu et al. have made significant advancements in the field of photonic crystal surface-emitting lasers (PCSELs) by ingeniously adjusting the photonic crystal structure to achieve linear dispersion of the accidental Dirac point around the $\Gamma$ point. This breakthrough increases the mode spacing by orders of magnitude and eliminates distributed in-plane feedback, enabling the development of larger-area single-mode PCSELs with substantially higher output power\cite{chua_larger-area_2014,bravo-abad_enabling_2012}.
Subsequently, Contractor et al. experimentally demonstrated a scalable BerkSEL via open-Dirac singularities\cite{C.N.R2022}. Remarkably, these lasers sustain large-area cavity modes as the cavity size increases. However, cavity modes arise from gapless open-Dirac points and lack topological bandgap protection. Besides, these cavity modes emerge from a single kind of lattice, and the corresponding mode area is always locked to the structure size, leading to a lack of flexibility for modulation and manipulation.  

On the other hand, topological phases have garnered substantial attention due to their robustness against lattice disorders\cite{wangReflectionFree2008,yangExperimental2013,khanikaevPhotonic2013,fangRealizing2012,rechtsmanPhotonic2013,hePhotonic2016,maGuiding2015,chengRobust2016,zhuTopological2018,barikTwodimensionally2016,wuScheme2015,xuAccidental2016,wangThreedimensional2016,luSymmetryprotected2016,gaoTopologically2018,shalaevRobust2019}. For example, chiral symmetric zero-energy topological boundary modes have been found in the one-dimensional (1D) SSH model\cite{ryuTopological2002}. Moreover, higher-order topology in two and higher dimensions emerges with lower-dimensional topological boundary modes such as zero-dimensional (0D) corner modes\cite{xieSecondorder2018,xieHigherorder2020,xieVisualization2019,xieHigherorder2021} and 1D hinge modes\cite{benalcazarQuantized2017,langbehnReflectionSymmetric2017,songDimensional2017,benalcazarElectric2017}. In experiments, these topological modes have been realized in different platforms ranging from electronic materials\cite{zhangCatalogue2019,fangAnomalous2003} to classical waves\cite{wangReflectionFree2008,yangExperimental2013,khanikaevPhotonic2013,fangRealizing2012}. These multidimensional topological boundary modes sustain wide applications such as optical communications\cite{shalaevRobust2019,yangTerahertz2020}, topological lasers\cite{zengElectrically2020,Ota.K.W.I.A2018,bandresTopological2018}, and quantum computing\cite{miyakeQuantum2010,elseSymmetryProtected2012}.

Among these studies, the localized topological boundary modes typically emerge from a single lattice (with other lattices serving as the gapped boundaries)\cite{wangReflectionFree2008,yangExperimental2013,khanikaevPhotonic2013,fangRealizing2012,rechtsmanPhotonic2013}. Recently, the novel topological modes that emerged from topological heterostructure have been studied, showing unconventional characteristics of modes\cite{wangValleylocked2020,wangTopological2021a,chenPhotonic2021,wang2024TI}. For example, combining 1D topological edge modes and the materials with Dirac points results in unprecedented two-dimensional (2D) large-area waveguide modes with robustness against disorders. However, albeit possessing large area properties, these 1D large-area topological edge modes are densely distributed across the energy domain, resulting in the lack of spectral-isolated characters. 

\begin{figure}[h]%
\centering
\includegraphics[width =\columnwidth]{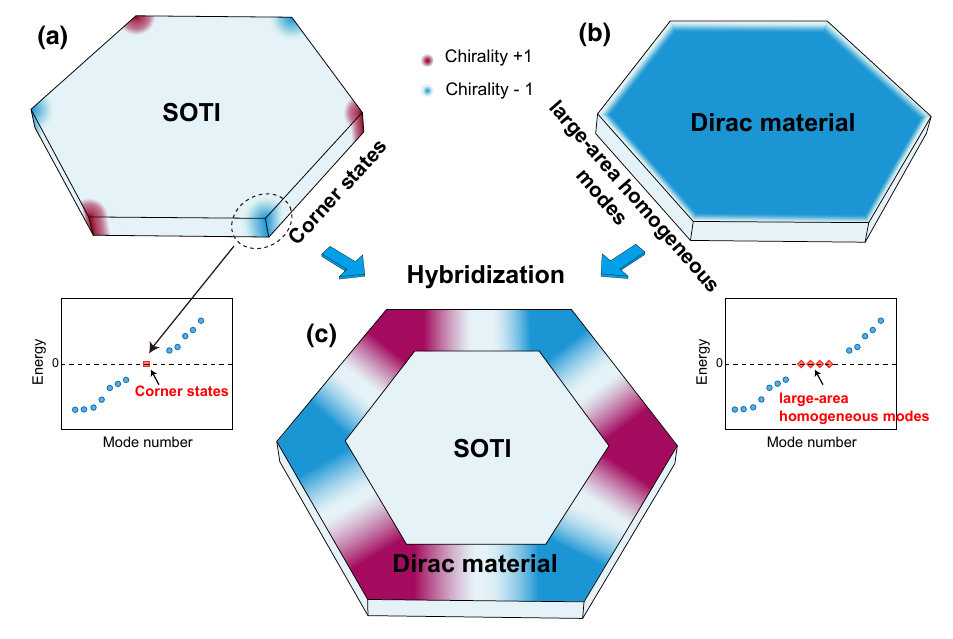}
\caption{\textbf{(a)}, Schematic diagram of corner states with different chirality in a SOTI, which have been labeled by the blue and red colors. A dashed line marks one corner structure of the $C_6$ symmetric sample and corresponding eigenvalues are shown in the inset, where the red dot represents the eigenvalue of the corner state. \textbf{(b)}, Schematics of the LHMs of DM. The inset shows the eigenvalues of the cavity. \textbf{(c)}, The large-area topological corner mode.}
\label{fig1}
\end{figure}

In this study, we theoretically propose and experimentally demonstrate a spectral-isolated photonic topological corner mode with a tunable mode area and stable frequency through hybridizing a second-order topological insulator (SOTI) with a gapless Dirac material (DM)\cite{wuScheme2015}. By matching the eigenfrequency of 0D topological corner modes in SOTI and large-area homogeneous mode (LHM) in DM, we successfully realized a 2D TLCM originating from the hybridization between the topological corner modes and LHM of DM. We observe that the frequency of TLCM remains unchanged, and the spectral-isolated character is well-preserved when the mode area changes. Moreover, we can control the mode area by changing the couplings' difference in the inner topological non-trivial lattice. Finally, we demonstrate that the TLCM is topologically protected and robust against lattice disorders inside the device. TLCM offers promising opportunities in various fields due to its mode-area-tunable and frequency-stable properties. This TLCM can improve the energy efficiency and beam quality of laser systems\cite{gaoDiracvortex2020,C.N.R2022}. Moreover, our TLCM can also be applied to facilitate high-radiance, single-frequency lasing\cite{zhao_non-hermitian_2019}. In optical communication, this mode enables channels with reduced signal loss and increased resilience to noise, paving the way for high-speed networks\cite{bahari_photonic_2021}.}

{\it{Theory and numerics}}.---{We begin with introducing the physical mechanism behind the formation of TLCM. In a SOTI featuring $C_6$ symmetry and described by a TBM, corner modes emerge and are exponentially localized at each corner, as depicted in Fig. 1(a). We plot the eigenvalues for one corner structure in the inset below for simplicity. This inset demonstrates the presence of in-gap corner modes, represented by the red dot (detailed discussion on topological invariants can be found in Appendix E). Subsequently, when examining a finite-size photonic crystal with a Dirac point at the center of the Brillouin zone, we can observe four LHMs, indicated by the red dots in Fig. 1(b) (a detailed discussion on LHMs can be found in Appendix B and C). The LHMs originate from the linear dispersion around the gapless double-Dirac cone in the band structure. In real space, these LHMs at the degenerate point correspond to a constant wave function, leading to a homogeneous distribution over all lattice sites (a more detailed discussion about the boundary conditions' effect on LHMs can be found in Appendix D). Subsequently, by joining the DM with the SOTI, as illustrated in Fig. 1(c), the corner modes hybridize with the LHM, yielding spectral-isolated photonic TLCM with a tunable mode area and stable frequency, as depicted in Fig. 1(c). Interestingly, this TLCM inherits the chirality of the corner modes in the SOTI and possesses both zero-energy character and non-zero wavefunction amplitude only at certain sub-lattices, making them significantly distinct from the LHMs in a DM.

\begin{figure}
\centering
\includegraphics[width =\columnwidth]{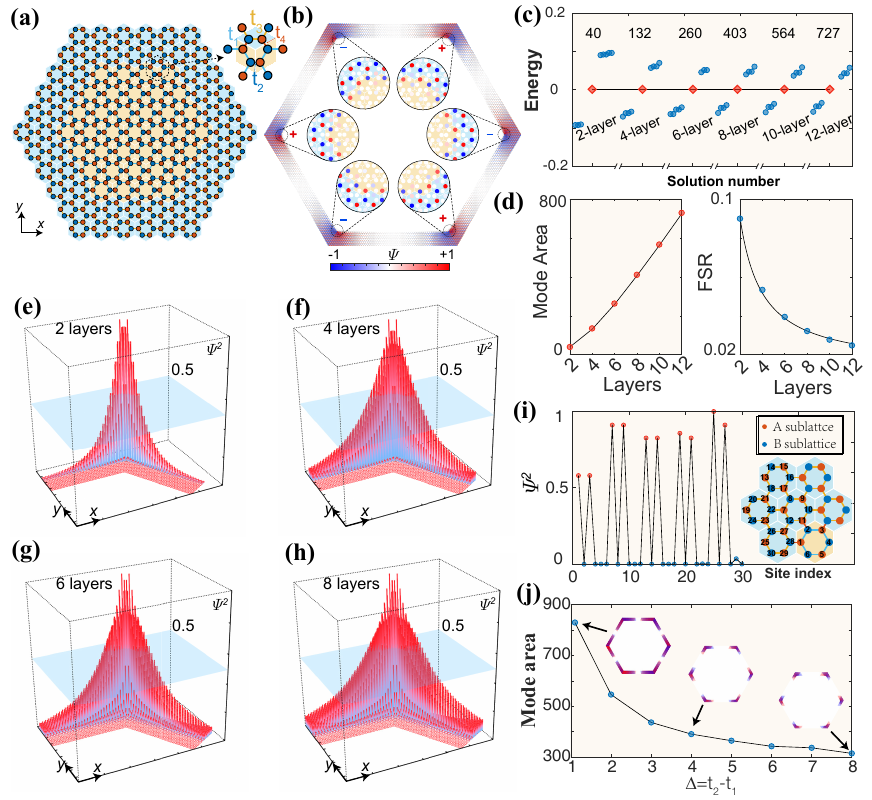}
\caption{\textbf{(a)}, Illustration of the TBM with $t_1=1,t_2=4,t_3=2,t_4=3$. \textbf{(b)}, Distribution of eigenmode in a finite-size structure, comprising of $60$ layers of SOTI unit cells and $8$ layers of gapless lattices. The chiral eigenvalues are indicated by '+' and '-'. \textbf{(c)}, Eigenvalues with different layers of gapless lattices as a function of mode area, where blue dots and red dots represent corner modes and other modes, respectively. \textbf{(d)}, FSR and mode area as functions of layers. \textbf{(e)-(h)}, TLCMs with $2$-layer, $4$-layer, $6$-layer, and $8$-layer gapless lattices. We only display $1/6$ of the entire structure due to the $C_6$ symmetry. The cyan plane indicates $\psi^2=max(\psi^2)/2$. \textbf{(i)}, Detailed TLCM around a corner, where the diminishing of the wavefunction amplitude at certain sublattice can be seen in the inset. \textbf{(j)}, Mode area as a function of $\Delta$. The insets show the $\psi$ distribution with $\Delta = 0.1,3,$ and $7$.} \label{fig2}
\end{figure}

To prove the existence of TLCM, we first consider a TBM as illustrated in Fig. 2(a) (see detailed discussion on building TBM in Appendix B). The TBM comprises non-trivial and gapless lattices, colored with yellow and blue backgrounds, respectively. Remarkably, the TBM possesses chiral (sublattice) symmetry for two kinds of lattices. In our design, four different couplings are denoted as $t_1=1,t_2=4,t_3=2,$ and $t_4=3$, respectively (inset of Fig. 1(a)). According to the schematic depicted in Fig. 1(a), we can obtain the Hamiltonian and calculate its eigenmodes, as shown in Fig. 2(b). We find that the eigenmodes are distributed over a large area around the corner structures, thus showing the existence of the TLCM. The insets in Fig. 2(b) display enlarged eigenmodes localized at corners, each possessing distinct chirality, labeled by “+” and “-”\cite{nohTopological2018}.

\begin{figure}
\centering
\includegraphics[width =\columnwidth]{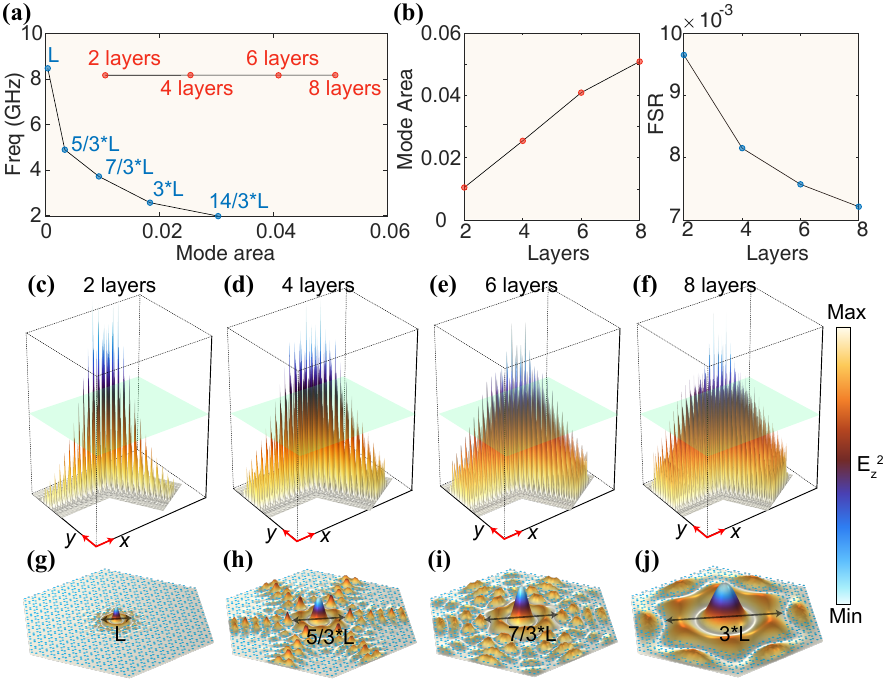}
\caption{Full-wave simulation results of the photonic TLCM and photonic crystal defect cavity. \textbf{(a)}, Eigenfrequencies of the photonic crystal defect cavity (blue line) exhibit a decay as the mode area increases, while the eigenfrequencies of the corner modes (red) remain nearly constant. \textbf{(b)}, Depicts the relationship between layers, FSR, and mode area, consistent with results obtained from the TBM. \textbf{(c)-(f)}, The distribution of $\left| E_z \right|^2$ of TLCMs with various layers. \textbf{(g)-(h)}, Display the distribution of $\left| E_z \right|^2$ for cavities of various sizes, with the blue dots representing the structure of photonic crystals and $L=2a$.}\label{fig3}
\end{figure}

Next, we study the relation between the eigenvalue, mode area, and the free spectral range (FSR) of the TLCM. In our work, the FSR is defined as $FSR={{( E_b-E_0 )}/{E_0}}$, where $E_0$ represents the eigenvalue of the TLCM and $E_b$ represents the eigenvalue of the nearest mode to TLCM. The mode area is defined as $M_a=\sum\nolimits_{n=1}^N{\max ( \left| \psi_n \right|^2-\frac{\left| \psi _m \right|^2}{2},0 )}$, where $\psi _m=\max ( \psi )$, $n$ is index of sites, and $\psi_n$ is the $\psi$ on the site labeled $n$. Figure. 2(c) displays nine eigenvalues closest to zero energy of the TLCM with different mode areas. We emphasize that the red dots represent the average of six TLCMs near zero energy. According to the eigenvalues, we can calculate the mode area and FSR trends as the number of layers changes in Fig. 2(d). It should be noted that the layers here indicate gapless lattice layers (or regions of DM), not referring to the entire structure. As the layers increase, the FSR decreases proportionally to $1/x$ while the mode area increases proportionally to $x$. To demonstrate the mode-area-tunable character, we visualize these distributions for structures with 2 to 8 gapless lattice layers in Fig. 2(e)-2(h). We can observe the increase in mode area when the layers of DM increase. The dashed blue planes in these figures indicate half the height of the $\psi^2$. Here, we only display $1/6$ of the entire eigenvector since the $C_6$ symmetry of our structure. 

Compared to the LHMs in gapless lattices, our TLCM exhibits distinct chiralities inherited from the chiralities of the corner modes in the topological non-trivial lattices, which is a highly non-trivial result. The chiral symmetry, i.e., the sublattice symmetry, restrains the eigenvalues to be symmetric concerning the zero-energy level. Therefore, the emerging topological corner modes and the TLCM are all pinned at zero energy, which may have chirality +1 or -1 eigenvalues concerning the chiral operators. This unique feature will maximize the FSR\cite{deng_observation_2022} of the TLCM (under the same bandgap size) and hence lead to better topological protection. Even in the lattices in DM that are far from the corner structure, the wave profile possesses a similar spatial distribution to those in the corner SOTI. Namely, there is only a non-zero wavefunction at one sublattice, while for other sublattice sites, the wavefunction is precisely zero, which is clear evidence of the chiral symmetry protection (see Fig. 2(i)).

Besides the chiralities, the TLCMs possess another fundamental difference from the LHMs. In a single SOTI with open boundary condition, when the mass term is increased (decreased) by adjusting the difference in the intercell and intracell couplings ($\Delta = t_2-t_1$), the decay length of corner modes becomes smaller (larger). When we combine the SOTI with DM, the resulting TLCM inherits the decay behavior of the original corner modes. Consequently, in the combined structure, as shown in Fig. 2(j), when the mass term in the SOTI structure is tuned, the TLCMs' mode area is also changed. This unique physical property of TLCMs enables the potential active control of mode area by applying the nonlinearity in the lattice. 

Additionally, the mode area is essential for the beam collimation of laser. The Fourier transform of a large-area mode is concentrated around specific $k$ components, whereas the Fourier transform of a small-area mode is relatively evenly distributed across many $k$ components. Therefore, this concentration of $k$ components in large-area modes contributes to their smaller beam divergence angles, enhancing the overall beam quality.}

{\it{Photonic crystals design}}.---{The TLCM is not limited to the theoretical TBM but can be realized in photonic crystals (more discussions about the bands structure and topological invariants can be found in Appendix C and E). Here, we design a photonic TLCM through full-wave simulations. We focus on constructing $1/6$ of the $C_6$ structures in Fig. 1(a). We create a photonic crystal slab consisting of three layers of SOTI, two or four layers of gapless lattices, and three layers of ordinary insulator (OI) lattices. The eigenfrequencies and eigenfields of TLCMs with varying layers of gapless lattices are presented in Fig. 3(a) and 3(c)-3(f). We observe that the frequencies of the photonic TLCM remain nearly unchanged as the number of layers increases. This unique behavior demonstrates the mode-area-tunable and frequency-stable characters of the photonic TLCM. We also design a photonic crystal defect cavity by removing several dielectric rods in the center of photonic crystals for comparison, as shown in Fig. 3(g)-(j). The frequency of the defective mode changes drastically as the system size increases as depicted in Fig. 3(a). We also calculate the relation between the FSR and mode area as a function of different numbers of layers and plot it in Fig. 3(b). We find that the mode area and FSR trend as the number of layers and the field distribution are similar to the TBM, which can demonstrate mode-area-tunable and frequency-stable characters of the photonic TLCM.}

{\it{Experimental realizations}}.---{To experimentally observe photonic TLCM, we fabricated two samples with 2-layer and 4-layer gapless latices, as illustrated in Fig. 4(a), 4(b), and 4(e), respectively (Detailed experimental procedures and the fabrication of samples can be found in Appendix A). The presence of the TLCMs is clearly shown from our near-field scanning results (Fig. 4(c) and 4(d)), and as the number of layers increases, the mode area gradually expands while maintaining a nearly constant frequency. Specifically, we observe the TLCM at 8.205 GHz and 8.200 GHz for 2-layer and 4-layer gapless latices, respectively. To demonstrate the spectral character of the photonic TLCM, we present transmission spectra in Fig. 4(f), which reveals a peak around 8.2 GHz, corresponding to the frequency of photonic TLCM. 

\begin{figure}
\centering
\includegraphics[width =\columnwidth]{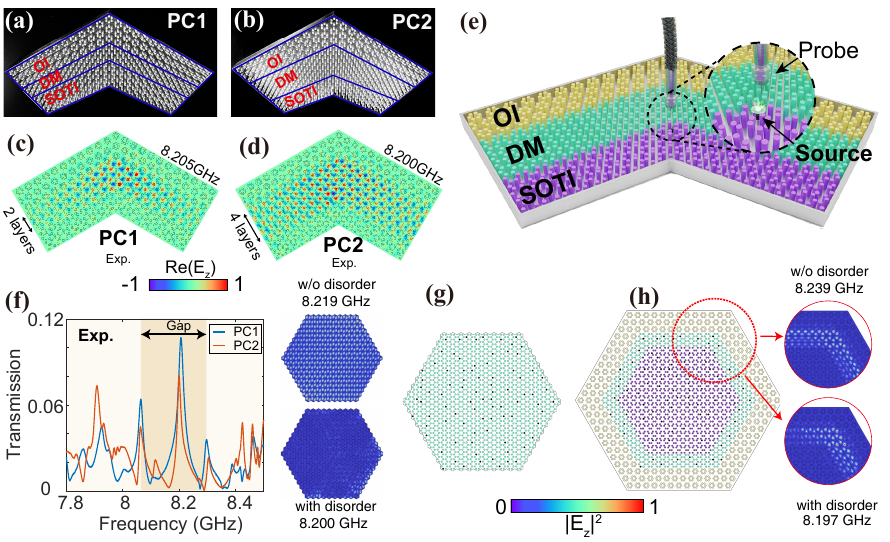}
 \caption{Experimental realizations.
 \textbf{(a)} and \textbf{(b)}, Photographs of PCs with two layers of DM lattices and four layers of DM lattices, labeled as PC1 and PC2. \textbf{(c)} and \textbf{(d)}, 
Measured electric field ($E_z$) distributions in different PCs. \textbf{(e)}, The schematics of the experimental setup used for near-field scanning and transmission spectra measurements. \textbf{(f)}, The transmission spectra for the three types of PCs. Resonance peaks correspond to TLCM and exhibit single-mode characteristics. The blue and orange solid lines represent PC1 and PC2, respectively, with the yellow shadow marking the frequency range between the TLCM and adjacent other modes. \textbf{(g)}, DM cavity structure and corresponding $E_{z}^{2}$ distribution with and without disorder. The black dots represent the position of dielectric pillars that are removed. The top(bottom) right inset plots the $E_{z}^{2}$ distribution without(with) disorder. \textbf{(h)}, SOTI/DM/OI structure and corresponding $E_{z}^{2}$ field with and without disorder. Here, we only show the field in a red dashed frame. The top(bottom) right inset plots the $E_{z}^{2}$ field without(with) disorder.}\label{fig4}
\end{figure}

To verify photonic TLCM's topological robustness, we simulated the eigenmodes of two structures, as shown in Fig. 4(g) and 4(h). The first structure consists of ten layers of gapless lattices, and the second structure is a photonic heterostructure consisting of sxi layers of SOTI, four layers of DM, and four layers of OI. Then, we introduce disorders by randomly removing dielectric pillars in the photonic heterostructure to observe whether the TLCM is preserved. We use black dots to mark the location of pillars that have been removed in Fig. 4(g) and 4(h). Under the same kind of lattice disorder, the LHM in the gapless photonic crystal is destroyed as there is no homogeneously distributed field (see other field distribution in Appendix C). At the same time, the TLCMs can still be identified with a frequency that remains almost unchanged. This observation reveals the topological protection of our TLCM compared to the LHM in the gapless photonic crystal\cite{C.N.R2022}.}

{\it{Discussion}}.---{This study proposes a spectral-isolated photonic topological corner mode with a tunable mode area and stable frequency. Our results demonstrate the unique physical properties of the emergent collective modes in the hybridized lattices, which consist of higher-order topology and DM. The observed TLCM holds promise for integrated photonic devices\cite{wangValleylocked2020,chenPhotonic2021} and topological surface-emitting lasers with high power and robustness\cite{chenHigh2017,liuPeriodic2016,eatonLasing2016}. We expect further explorations of exotic collective modes arising from hybridized lattices such as three-dimensional large-volume topological resonators \cite{wang2024TI}. We also expect the large area non-reciprocal modes to arise from combining the time-reversal breaking system with the DM\cite{wangTopological2021a}. Finally, our observed phenomena are not limited to the TB model and photonic crystals and may readily be realized in many other classical waves, such as acoustics \cite{wangValleylocked2020}, mechanics\cite{ma2019topological}, plasmonics\cite{politano2017optoelectronic}, etc.}

\begin{acknowledgments}
We acknowledge helpful discussions with Feng Liu. This work was financially supported by the National Key R$\&$D Program of China (Grants No. 2022YFA1404300, and 2021YFA1401103), the New Cornerstone Science Foundation, Hong Kong Research Grant Council (AoE/P-701/20, 17309021), National Natural Science Foundation of China (Grants No. 12174189 and 11834007), Stable Support Program for Higher Education Institutions of Shenzhen (No.20220817185604001) and the startup funding of the Chinese University of Hong Kong, Shenzhen (UDF01002563).
\end{acknowledgments}

 \bibliography{bib_large}

\begin{thebibliography}{57}%
\makeatletter
\providecommand \@ifxundefined [1]{%
 \@ifx{#1\undefined}
}%
\providecommand \@ifnum [1]{%
 \ifnum #1\expandafter \@firstoftwo
 \else \expandafter \@secondoftwo
 \fi
}%
\providecommand \@ifx [1]{%
 \ifx #1\expandafter \@firstoftwo
 \else \expandafter \@secondoftwo
 \fi
}%
\providecommand \natexlab [1]{#1}%
\providecommand \enquote  [1]{``#1''}%
\providecommand \bibnamefont  [1]{#1}%
\providecommand \bibfnamefont [1]{#1}%
\providecommand \citenamefont [1]{#1}%
\providecommand \href@noop [0]{\@secondoftwo}%
\providecommand \href [0]{\begingroup \@sanitize@url \@href}%
\providecommand \@href[1]{\@@startlink{#1}\@@href}%
\providecommand \@@href[1]{\endgroup#1\@@endlink}%
\providecommand \@sanitize@url [0]{\catcode `\\12\catcode `\$12\catcode `\&12\catcode `\#12\catcode `\^12\catcode `\_12\catcode `\%12\relax}%
\providecommand \@@startlink[1]{}%
\providecommand \@@endlink[0]{}%
\providecommand \url  [0]{\begingroup\@sanitize@url \@url }%
\providecommand \@url [1]{\endgroup\@href {#1}{\urlprefix }}%
\providecommand \urlprefix  [0]{URL }%
\providecommand \Eprint [0]{\href }%
\providecommand \doibase [0]{https://doi.org/}%
\providecommand \selectlanguage [0]{\@gobble}%
\providecommand \bibinfo  [0]{\@secondoftwo}%
\providecommand \bibfield  [0]{\@secondoftwo}%
\providecommand \translation [1]{[#1]}%
\providecommand \BibitemOpen [0]{}%
\providecommand \bibitemStop [0]{}%
\providecommand \bibitemNoStop [0]{.\EOS\space}%
\providecommand \EOS [0]{\spacefactor3000\relax}%
\providecommand \BibitemShut  [1]{\csname bibitem#1\endcsname}%
\let\auto@bib@innerbib\@empty
\bibitem [{\citenamefont {Chen}\ \emph {et~al.}(2017)\citenamefont {Chen}, \citenamefont {Zhang}, \citenamefont {Lee}, \citenamefont {Han},\ and\ \citenamefont {Nurmikko}}]{chenHigh2017}%
  \BibitemOpen
  \bibfield  {author} {\bibinfo {author} {\bibfnamefont {S.}~\bibnamefont {Chen}}, \bibinfo {author} {\bibfnamefont {C.}~\bibnamefont {Zhang}}, \bibinfo {author} {\bibfnamefont {J.}~\bibnamefont {Lee}}, \bibinfo {author} {\bibfnamefont {J.}~\bibnamefont {Han}},\ and\ \bibinfo {author} {\bibfnamefont {A.}~\bibnamefont {Nurmikko}},\ }\bibfield  {title} {\bibinfo {title} {High- {{{\emph{Q}}}} , {{Low}}-{{Threshold Monolithic Perovskite Thin}}-{{Film Vertical}}-{{Cavity Lasers}}},\ }\href {https://doi.org/10.1002/adma.201604781} {\bibfield  {journal} {\bibinfo  {journal} {Adv. Mater.}\ }\textbf {\bibinfo {volume} {29}},\ \bibinfo {pages} {1604781} (\bibinfo {year} {2017})}\BibitemShut {NoStop}%
\bibitem [{\citenamefont {Liu}\ \emph {et~al.}(2016)\citenamefont {Liu}, \citenamefont {Niu}, \citenamefont {Wu}, \citenamefont {Cong}, \citenamefont {Wang}, \citenamefont {Zeng}, \citenamefont {He}, \citenamefont {Fu}, \citenamefont {Fu}, \citenamefont {Yu}, \citenamefont {Jin}, \citenamefont {Liu},\ and\ \citenamefont {Sum}}]{liuPeriodic2016}%
  \BibitemOpen
  \bibfield  {author} {\bibinfo {author} {\bibfnamefont {X.}~\bibnamefont {Liu}}, \bibinfo {author} {\bibfnamefont {L.}~\bibnamefont {Niu}}, \bibinfo {author} {\bibfnamefont {C.}~\bibnamefont {Wu}}, \bibinfo {author} {\bibfnamefont {C.}~\bibnamefont {Cong}}, \bibinfo {author} {\bibfnamefont {H.}~\bibnamefont {Wang}}, \bibinfo {author} {\bibfnamefont {Q.}~\bibnamefont {Zeng}}, \bibinfo {author} {\bibfnamefont {H.}~\bibnamefont {He}}, \bibinfo {author} {\bibfnamefont {Q.}~\bibnamefont {Fu}}, \bibinfo {author} {\bibfnamefont {W.}~\bibnamefont {Fu}}, \bibinfo {author} {\bibfnamefont {T.}~\bibnamefont {Yu}}, \bibinfo {author} {\bibfnamefont {C.}~\bibnamefont {Jin}}, \bibinfo {author} {\bibfnamefont {Z.}~\bibnamefont {Liu}},\ and\ \bibinfo {author} {\bibfnamefont {T.~C.}\ \bibnamefont {Sum}},\ }\bibfield  {title} {\bibinfo {title} {Periodic {{Organic}}\textendash{{Inorganic Halide Perovskite Microplatelet Arrays}} on {{Silicon Substrates}} for {{Room}}-{{Temperature Lasing}}},\ }\href
  {https://doi.org/10.1002/advs.201600137} {\bibfield  {journal} {\bibinfo  {journal} {Adv. Sci}\ }\textbf {\bibinfo {volume} {3}},\ \bibinfo {pages} {1600137} (\bibinfo {year} {2016})}\BibitemShut {NoStop}%
\bibitem [{\citenamefont {Eaton}\ \emph {et~al.}(2016{\natexlab{a}})\citenamefont {Eaton}, \citenamefont {Lai}, \citenamefont {Gibson}, \citenamefont {Wong}, \citenamefont {Dou}, \citenamefont {Ma}, \citenamefont {Wang}, \citenamefont {Leone},\ and\ \citenamefont {Yang}}]{eatonLasing2016}%
  \BibitemOpen
  \bibfield  {author} {\bibinfo {author} {\bibfnamefont {S.~W.}\ \bibnamefont {Eaton}}, \bibinfo {author} {\bibfnamefont {M.}~\bibnamefont {Lai}}, \bibinfo {author} {\bibfnamefont {N.~A.}\ \bibnamefont {Gibson}}, \bibinfo {author} {\bibfnamefont {A.~B.}\ \bibnamefont {Wong}}, \bibinfo {author} {\bibfnamefont {L.}~\bibnamefont {Dou}}, \bibinfo {author} {\bibfnamefont {J.}~\bibnamefont {Ma}}, \bibinfo {author} {\bibfnamefont {L.-W.}\ \bibnamefont {Wang}}, \bibinfo {author} {\bibfnamefont {S.~R.}\ \bibnamefont {Leone}},\ and\ \bibinfo {author} {\bibfnamefont {P.}~\bibnamefont {Yang}},\ }\bibfield  {title} {\bibinfo {title} {Lasing in robust cesium lead halide perovskite nanowires},\ }\href {https://doi.org/10.1073/pnas.1600789113} {\bibfield  {journal} {\bibinfo  {journal} {PNAS}\ }\textbf {\bibinfo {volume} {113}},\ \bibinfo {pages} {1993} (\bibinfo {year} {2016}{\natexlab{a}})}\BibitemShut {NoStop}%
\bibitem [{\citenamefont {Eaton}\ \emph {et~al.}(2016{\natexlab{b}})\citenamefont {Eaton}, \citenamefont {Fu}, \citenamefont {Wong}, \citenamefont {Ning},\ and\ \citenamefont {Yang}}]{eatonSemiconductor2016}%
  \BibitemOpen
  \bibfield  {author} {\bibinfo {author} {\bibfnamefont {S.~W.}\ \bibnamefont {Eaton}}, \bibinfo {author} {\bibfnamefont {A.}~\bibnamefont {Fu}}, \bibinfo {author} {\bibfnamefont {A.~B.}\ \bibnamefont {Wong}}, \bibinfo {author} {\bibfnamefont {C.-Z.}\ \bibnamefont {Ning}},\ and\ \bibinfo {author} {\bibfnamefont {P.}~\bibnamefont {Yang}},\ }\bibfield  {title} {\bibinfo {title} {Semiconductor nanowire lasers},\ }\href {https://doi.org/10.1038/natrevmats.2016.28} {\bibfield  {journal} {\bibinfo  {journal} {Nat. Rev. Mater.}\ }\textbf {\bibinfo {volume} {1}},\ \bibinfo {pages} {16028} (\bibinfo {year} {2016}{\natexlab{b}})}\BibitemShut {NoStop}%
\bibitem [{\citenamefont {Gao}\ \emph {et~al.}(2020)\citenamefont {Gao}, \citenamefont {Yang}, \citenamefont {Lin}, \citenamefont {Zhang}, \citenamefont {Li}, \citenamefont {Bo}, \citenamefont {Wang},\ and\ \citenamefont {Lu}}]{gaoDiracvortex2020}%
  \BibitemOpen
  \bibfield  {author} {\bibinfo {author} {\bibfnamefont {X.}~\bibnamefont {Gao}}, \bibinfo {author} {\bibfnamefont {L.}~\bibnamefont {Yang}}, \bibinfo {author} {\bibfnamefont {H.}~\bibnamefont {Lin}}, \bibinfo {author} {\bibfnamefont {L.}~\bibnamefont {Zhang}}, \bibinfo {author} {\bibfnamefont {J.}~\bibnamefont {Li}}, \bibinfo {author} {\bibfnamefont {F.}~\bibnamefont {Bo}}, \bibinfo {author} {\bibfnamefont {Z.}~\bibnamefont {Wang}},\ and\ \bibinfo {author} {\bibfnamefont {L.}~\bibnamefont {Lu}},\ }\bibfield  {title} {\bibinfo {title} {Dirac-vortex topological cavities},\ }\href {https://doi.org/10.1038/s41565-020-0773-7} {\bibfield  {journal} {\bibinfo  {journal} {Nat. Nanotechnol}\ }\textbf {\bibinfo {volume} {15}},\ \bibinfo {pages} {1012} (\bibinfo {year} {2020})}\BibitemShut {NoStop}%
\bibitem [{\citenamefont {Schawlow}\ and\ \citenamefont {Townes}(1958)}]{schawlowInfrared1958}%
  \BibitemOpen
  \bibfield  {author} {\bibinfo {author} {\bibfnamefont {A.~L.}\ \bibnamefont {Schawlow}}\ and\ \bibinfo {author} {\bibfnamefont {C.~H.}\ \bibnamefont {Townes}},\ }\bibfield  {title} {\bibinfo {title} {Infrared and {{Optical Masers}}},\ }\href {https://doi.org/10.1103/PhysRev.112.1940} {\bibfield  {journal} {\bibinfo  {journal} {Phys. Rev.}\ }\textbf {\bibinfo {volume} {112}},\ \bibinfo {pages} {1940} (\bibinfo {year} {1958})}\BibitemShut {NoStop}%
\bibitem [{\citenamefont {Kogelnik}\ and\ \citenamefont {Shank}(1971)}]{kogelnikSTIMULATED1971}%
  \BibitemOpen
  \bibfield  {author} {\bibinfo {author} {\bibfnamefont {H.}~\bibnamefont {Kogelnik}}\ and\ \bibinfo {author} {\bibfnamefont {C.~V.}\ \bibnamefont {Shank}},\ }\bibfield  {title} {\bibinfo {title} {{{STIMULATED EMISSION IN A PERIODIC STRUCTURE}}},\ }\href {https://doi.org/10.1063/1.1653605} {\bibfield  {journal} {\bibinfo  {journal} {Appl. Phys. Lett.}\ }\textbf {\bibinfo {volume} {18}},\ \bibinfo {pages} {152} (\bibinfo {year} {1971})}\BibitemShut {NoStop}%
\bibitem [{\citenamefont {Soda}\ \emph {et~al.}(1979)\citenamefont {Soda}, \citenamefont {Iga}, \citenamefont {Kitahara},\ and\ \citenamefont {Suematsu}}]{sodaGaInAsP1979}%
  \BibitemOpen
  \bibfield  {author} {\bibinfo {author} {\bibfnamefont {H.}~\bibnamefont {Soda}}, \bibinfo {author} {\bibfnamefont {K.-i.}\ \bibnamefont {Iga}}, \bibinfo {author} {\bibfnamefont {C.}~\bibnamefont {Kitahara}},\ and\ \bibinfo {author} {\bibfnamefont {Y.}~\bibnamefont {Suematsu}},\ }\bibfield  {title} {\bibinfo {title} {{{GaInAsP}}/{{InP Surface Emitting Injection Lasers}}},\ }\href {https://doi.org/10.1143/JJAP.18.2329} {\bibfield  {journal} {\bibinfo  {journal} {Jpn. J. Appl. Phys.}\ }\textbf {\bibinfo {volume} {18}},\ \bibinfo {pages} {2329} (\bibinfo {year} {1979})}\BibitemShut {NoStop}%
\bibitem [{\citenamefont {Imada}\ \emph {et~al.}(1999)\citenamefont {Imada}, \citenamefont {Noda}, \citenamefont {Chutinan}, \citenamefont {Tokuda}, \citenamefont {Murata},\ and\ \citenamefont {Sasaki}}]{imadaCoherent1999}%
  \BibitemOpen
  \bibfield  {author} {\bibinfo {author} {\bibfnamefont {M.}~\bibnamefont {Imada}}, \bibinfo {author} {\bibfnamefont {S.}~\bibnamefont {Noda}}, \bibinfo {author} {\bibfnamefont {A.}~\bibnamefont {Chutinan}}, \bibinfo {author} {\bibfnamefont {T.}~\bibnamefont {Tokuda}}, \bibinfo {author} {\bibfnamefont {M.}~\bibnamefont {Murata}},\ and\ \bibinfo {author} {\bibfnamefont {G.}~\bibnamefont {Sasaki}},\ }\bibfield  {title} {\bibinfo {title} {Coherent two-dimensional lasing action in surface-emitting laser with triangular-lattice photonic crystal structure},\ }\href {https://doi.org/10.1063/1.124361} {\bibfield  {journal} {\bibinfo  {journal} {Appl. Phys. Lett.}\ }\textbf {\bibinfo {volume} {75}},\ \bibinfo {pages} {316} (\bibinfo {year} {1999})}\BibitemShut {NoStop}%
\bibitem [{\citenamefont {Yoshida}\ \emph {et~al.}(2019)\citenamefont {Yoshida}, \citenamefont {De~Zoysa}, \citenamefont {Ishizaki}, \citenamefont {Tanaka}, \citenamefont {Kawasaki}, \citenamefont {Hatsuda}, \citenamefont {Song}, \citenamefont {Gelleta},\ and\ \citenamefont {Noda}}]{yoshidaDoublelattice2019}%
  \BibitemOpen
  \bibfield  {author} {\bibinfo {author} {\bibfnamefont {M.}~\bibnamefont {Yoshida}}, \bibinfo {author} {\bibfnamefont {M.}~\bibnamefont {De~Zoysa}}, \bibinfo {author} {\bibfnamefont {K.}~\bibnamefont {Ishizaki}}, \bibinfo {author} {\bibfnamefont {Y.}~\bibnamefont {Tanaka}}, \bibinfo {author} {\bibfnamefont {M.}~\bibnamefont {Kawasaki}}, \bibinfo {author} {\bibfnamefont {R.}~\bibnamefont {Hatsuda}}, \bibinfo {author} {\bibfnamefont {B.}~\bibnamefont {Song}}, \bibinfo {author} {\bibfnamefont {J.}~\bibnamefont {Gelleta}},\ and\ \bibinfo {author} {\bibfnamefont {S.}~\bibnamefont {Noda}},\ }\bibfield  {title} {\bibinfo {title} {Double-lattice photonic-crystal resonators enabling high-brightness semiconductor lasers with symmetric narrow-divergence beams},\ }\href {https://doi.org/10.1038/s41563-018-0242-y} {\bibfield  {journal} {\bibinfo  {journal} {Nat. Mater}\ }\textbf {\bibinfo {volume} {18}},\ \bibinfo {pages} {121} (\bibinfo {year} {2019})}\BibitemShut {NoStop}%
\bibitem [{\citenamefont {Li}\ \emph {et~al.}(2024)\citenamefont {Li}, \citenamefont {Ma}, \citenamefont {Li}, \citenamefont {you}, \citenamefont {Liu}, \citenamefont {Yang}, \citenamefont {Xiang}, \citenamefont {Zhou},\ and\ \citenamefont {Zhang}}]{li2024CZM}%
  \BibitemOpen
  \bibfield  {author} {\bibinfo {author} {\bibfnamefont {Z.}~\bibnamefont {Li}}, \bibinfo {author} {\bibfnamefont {S.}~\bibnamefont {Ma}}, \bibinfo {author} {\bibfnamefont {S.}~\bibnamefont {Li}}, \bibinfo {author} {\bibfnamefont {O.}~\bibnamefont {you}}, \bibinfo {author} {\bibfnamefont {Y.}~\bibnamefont {Liu}}, \bibinfo {author} {\bibfnamefont {Q.}~\bibnamefont {Yang}}, \bibinfo {author} {\bibfnamefont {Y.}~\bibnamefont {Xiang}}, \bibinfo {author} {\bibfnamefont {P.}~\bibnamefont {Zhou}},\ and\ \bibinfo {author} {\bibfnamefont {S.}~\bibnamefont {Zhang}},\ }\href {https://arxiv.org/abs/2407.03390} {\bibinfo {title} {Observation of co-propagating chiral zero modes in magnetic photonic crystals}} (\bibinfo {year} {2024}),\ \Eprint {https://arxiv.org/abs/2407.03390} {arXiv:2407.03390 [cond-mat.mes-hall]} \BibitemShut {NoStop}%
\bibitem [{\citenamefont {Chua}\ \emph {et~al.}(2014)\citenamefont {Chua}, \citenamefont {Lu}, \citenamefont {Bravo-Abad}, \citenamefont {Joannopoulos},\ and\ \citenamefont {Soljačić}}]{chua_larger-area_2014}%
  \BibitemOpen
  \bibfield  {author} {\bibinfo {author} {\bibfnamefont {S.-L.}\ \bibnamefont {Chua}}, \bibinfo {author} {\bibfnamefont {L.}~\bibnamefont {Lu}}, \bibinfo {author} {\bibfnamefont {J.}~\bibnamefont {Bravo-Abad}}, \bibinfo {author} {\bibfnamefont {J.~D.}\ \bibnamefont {Joannopoulos}},\ and\ \bibinfo {author} {\bibfnamefont {M.}~\bibnamefont {Soljačić}},\ }\bibfield  {title} {\bibinfo {title} {Larger-area single-mode photonic crystal surface-emitting lasers enabled by an accidental dirac point},\ }\href {https://doi.org/10.1364/OL.39.002072} {\bibfield  {journal} {\bibinfo  {journal} {Opt. Lett.}\ }\textbf {\bibinfo {volume} {39}},\ \bibinfo {pages} {2072} (\bibinfo {year} {2014})},\ \bibinfo {note} {publisher: Optica Publishing Group}\BibitemShut {NoStop}%
\bibitem [{\citenamefont {Bravo-Abad}\ \emph {et~al.}(2012)\citenamefont {Bravo-Abad}, \citenamefont {Joannopoulos},\ and\ \citenamefont {Soljačić}}]{bravo-abad_enabling_2012}%
  \BibitemOpen
  \bibfield  {author} {\bibinfo {author} {\bibfnamefont {J.}~\bibnamefont {Bravo-Abad}}, \bibinfo {author} {\bibfnamefont {J.~D.}\ \bibnamefont {Joannopoulos}},\ and\ \bibinfo {author} {\bibfnamefont {M.}~\bibnamefont {Soljačić}},\ }\bibfield  {title} {\bibinfo {title} {Enabling single-mode behavior over large areas with photonic dirac cones},\ }\href {https://doi.org/10.1073/pnas.1207335109} {\bibfield  {journal} {\bibinfo  {journal} {Proceedings of the National Academy of Sciences}\ }\textbf {\bibinfo {volume} {109}},\ \bibinfo {pages} {9761} (\bibinfo {year} {2012})},\ \bibinfo {note} {publisher: Proceedings of the National Academy of Sciences}\BibitemShut {NoStop}%
\bibitem [{\citenamefont {Contractor}\ \emph {et~al.}(2022)\citenamefont {Contractor}, \citenamefont {Noh}, \citenamefont {Redjem}, \citenamefont {Qarony}, \citenamefont {Martin}, \citenamefont {Dhuey}, \citenamefont {Schwartzberg},\ and\ \citenamefont {Kant{\'e}}}]{C.N.R2022}%
  \BibitemOpen
  \bibfield  {author} {\bibinfo {author} {\bibfnamefont {R.}~\bibnamefont {Contractor}}, \bibinfo {author} {\bibfnamefont {W.}~\bibnamefont {Noh}}, \bibinfo {author} {\bibfnamefont {W.}~\bibnamefont {Redjem}}, \bibinfo {author} {\bibfnamefont {W.}~\bibnamefont {Qarony}}, \bibinfo {author} {\bibfnamefont {E.}~\bibnamefont {Martin}}, \bibinfo {author} {\bibfnamefont {S.}~\bibnamefont {Dhuey}}, \bibinfo {author} {\bibfnamefont {A.}~\bibnamefont {Schwartzberg}},\ and\ \bibinfo {author} {\bibfnamefont {B.}~\bibnamefont {Kant{\'e}}},\ }\bibfield  {title} {\bibinfo {title} {Scalable single-mode surface-emitting laser via open-{{Dirac}} singularities},\ }\href {https://doi.org/10.1038/s41586-022-05021-4} {\bibfield  {journal} {\bibinfo  {journal} {Nature}\ }\textbf {\bibinfo {volume} {608}},\ \bibinfo {pages} {692} (\bibinfo {year} {2022})}\BibitemShut {NoStop}%
\bibitem [{\citenamefont {Wang}\ \emph {et~al.}(2008)\citenamefont {Wang}, \citenamefont {Chong}, \citenamefont {Joannopoulos},\ and\ \citenamefont {Solja{\v c}i{\'c}}}]{wangReflectionFree2008}%
  \BibitemOpen
  \bibfield  {author} {\bibinfo {author} {\bibfnamefont {Z.}~\bibnamefont {Wang}}, \bibinfo {author} {\bibfnamefont {Y.~D.}\ \bibnamefont {Chong}}, \bibinfo {author} {\bibfnamefont {J.~D.}\ \bibnamefont {Joannopoulos}},\ and\ \bibinfo {author} {\bibfnamefont {M.}~\bibnamefont {Solja{\v c}i{\'c}}},\ }\bibfield  {title} {\bibinfo {title} {Reflection-{{Free One-Way Edge Modes}} in a {{Gyromagnetic Photonic Crystal}}},\ }\href {https://doi.org/10.1103/PhysRevLett.100.013905} {\bibfield  {journal} {\bibinfo  {journal} {Phys. Rev. Lett.}\ }\textbf {\bibinfo {volume} {100}},\ \bibinfo {pages} {013905} (\bibinfo {year} {2008})}\BibitemShut {NoStop}%
\bibitem [{\citenamefont {Yang}\ \emph {et~al.}(2013)\citenamefont {Yang}, \citenamefont {Poo}, \citenamefont {Wu}, \citenamefont {Gu},\ and\ \citenamefont {Chen}}]{yangExperimental2013}%
  \BibitemOpen
  \bibfield  {author} {\bibinfo {author} {\bibfnamefont {Y.}~\bibnamefont {Yang}}, \bibinfo {author} {\bibfnamefont {Y.}~\bibnamefont {Poo}}, \bibinfo {author} {\bibfnamefont {R.-x.}\ \bibnamefont {Wu}}, \bibinfo {author} {\bibfnamefont {Y.}~\bibnamefont {Gu}},\ and\ \bibinfo {author} {\bibfnamefont {P.}~\bibnamefont {Chen}},\ }\bibfield  {title} {\bibinfo {title} {Experimental demonstration of one-way slow wave in waveguide involving gyromagnetic photonic crystals},\ }\href {https://doi.org/10.1063/1.4809956} {\bibfield  {journal} {\bibinfo  {journal} {Appl. Phys. Lett.}\ }\textbf {\bibinfo {volume} {102}},\ \bibinfo {pages} {231113} (\bibinfo {year} {2013})}\BibitemShut {NoStop}%
\bibitem [{\citenamefont {Khanikaev}\ \emph {et~al.}(2013)\citenamefont {Khanikaev}, \citenamefont {Hossein~Mousavi}, \citenamefont {Tse}, \citenamefont {Kargarian}, \citenamefont {MacDonald},\ and\ \citenamefont {Shvets}}]{khanikaevPhotonic2013}%
  \BibitemOpen
  \bibfield  {author} {\bibinfo {author} {\bibfnamefont {A.~B.}\ \bibnamefont {Khanikaev}}, \bibinfo {author} {\bibfnamefont {S.}~\bibnamefont {Hossein~Mousavi}}, \bibinfo {author} {\bibfnamefont {W.-K.}\ \bibnamefont {Tse}}, \bibinfo {author} {\bibfnamefont {M.}~\bibnamefont {Kargarian}}, \bibinfo {author} {\bibfnamefont {A.~H.}\ \bibnamefont {MacDonald}},\ and\ \bibinfo {author} {\bibfnamefont {G.}~\bibnamefont {Shvets}},\ }\bibfield  {title} {\bibinfo {title} {Photonic topological insulators},\ }\href {https://doi.org/10.1038/nmat3520} {\bibfield  {journal} {\bibinfo  {journal} {Nat. Mater}\ }\textbf {\bibinfo {volume} {12}},\ \bibinfo {pages} {233} (\bibinfo {year} {2013})}\BibitemShut {NoStop}%
\bibitem [{\citenamefont {Fang}\ \emph {et~al.}(2012)\citenamefont {Fang}, \citenamefont {Yu},\ and\ \citenamefont {Fan}}]{fangRealizing2012}%
  \BibitemOpen
  \bibfield  {author} {\bibinfo {author} {\bibfnamefont {K.}~\bibnamefont {Fang}}, \bibinfo {author} {\bibfnamefont {Z.}~\bibnamefont {Yu}},\ and\ \bibinfo {author} {\bibfnamefont {S.}~\bibnamefont {Fan}},\ }\bibfield  {title} {\bibinfo {title} {Realizing effective magnetic field for photons by controlling the phase of dynamic modulation},\ }\href {https://doi.org/10.1038/nphoton.2012.236} {\bibfield  {journal} {\bibinfo  {journal} {Nat. Photonics}\ }\textbf {\bibinfo {volume} {6}},\ \bibinfo {pages} {782} (\bibinfo {year} {2012})}\BibitemShut {NoStop}%
\bibitem [{\citenamefont {Rechtsman}\ \emph {et~al.}(2013)\citenamefont {Rechtsman}, \citenamefont {Zeuner}, \citenamefont {Plotnik}, \citenamefont {Lumer}, \citenamefont {Podolsky}, \citenamefont {Dreisow}, \citenamefont {Nolte}, \citenamefont {Segev},\ and\ \citenamefont {Szameit}}]{rechtsmanPhotonic2013}%
  \BibitemOpen
  \bibfield  {author} {\bibinfo {author} {\bibfnamefont {M.~C.}\ \bibnamefont {Rechtsman}}, \bibinfo {author} {\bibfnamefont {J.~M.}\ \bibnamefont {Zeuner}}, \bibinfo {author} {\bibfnamefont {Y.}~\bibnamefont {Plotnik}}, \bibinfo {author} {\bibfnamefont {Y.}~\bibnamefont {Lumer}}, \bibinfo {author} {\bibfnamefont {D.}~\bibnamefont {Podolsky}}, \bibinfo {author} {\bibfnamefont {F.}~\bibnamefont {Dreisow}}, \bibinfo {author} {\bibfnamefont {S.}~\bibnamefont {Nolte}}, \bibinfo {author} {\bibfnamefont {M.}~\bibnamefont {Segev}},\ and\ \bibinfo {author} {\bibfnamefont {A.}~\bibnamefont {Szameit}},\ }\bibfield  {title} {\bibinfo {title} {Photonic {{Floquet}} topological insulators},\ }\href {https://doi.org/10.1038/nature12066} {\bibfield  {journal} {\bibinfo  {journal} {Nature}\ }\textbf {\bibinfo {volume} {496}},\ \bibinfo {pages} {196} (\bibinfo {year} {2013})}\BibitemShut {NoStop}%
\bibitem [{\citenamefont {He}\ \emph {et~al.}(2016)\citenamefont {He}, \citenamefont {Sun}, \citenamefont {Liu}, \citenamefont {Lu}, \citenamefont {Chen}, \citenamefont {Feng},\ and\ \citenamefont {Chen}}]{hePhotonic2016}%
  \BibitemOpen
  \bibfield  {author} {\bibinfo {author} {\bibfnamefont {C.}~\bibnamefont {He}}, \bibinfo {author} {\bibfnamefont {X.-C.}\ \bibnamefont {Sun}}, \bibinfo {author} {\bibfnamefont {X.-P.}\ \bibnamefont {Liu}}, \bibinfo {author} {\bibfnamefont {M.-H.}\ \bibnamefont {Lu}}, \bibinfo {author} {\bibfnamefont {Y.}~\bibnamefont {Chen}}, \bibinfo {author} {\bibfnamefont {L.}~\bibnamefont {Feng}},\ and\ \bibinfo {author} {\bibfnamefont {Y.-F.}\ \bibnamefont {Chen}},\ }\bibfield  {title} {\bibinfo {title} {Photonic topological insulator with broken time-reversal symmetry},\ }\href {https://doi.org/10.1073/pnas.1525502113} {\bibfield  {journal} {\bibinfo  {journal} {PNAS}\ }\textbf {\bibinfo {volume} {113}},\ \bibinfo {pages} {4924} (\bibinfo {year} {2016})}\BibitemShut {NoStop}%
\bibitem [{\citenamefont {Ma}\ \emph {et~al.}(2015)\citenamefont {Ma}, \citenamefont {Khanikaev}, \citenamefont {Mousavi},\ and\ \citenamefont {Shvets}}]{maGuiding2015}%
  \BibitemOpen
  \bibfield  {author} {\bibinfo {author} {\bibfnamefont {T.}~\bibnamefont {Ma}}, \bibinfo {author} {\bibfnamefont {A.~B.}\ \bibnamefont {Khanikaev}}, \bibinfo {author} {\bibfnamefont {S.~H.}\ \bibnamefont {Mousavi}},\ and\ \bibinfo {author} {\bibfnamefont {G.}~\bibnamefont {Shvets}},\ }\bibfield  {title} {\bibinfo {title} {Guiding {{Electromagnetic Waves}} around {{Sharp Corners}}: {{Topologically Protected Photonic Transport}} in {{Metawaveguides}}},\ }\href {https://doi.org/10.1103/PhysRevLett.114.127401} {\bibfield  {journal} {\bibinfo  {journal} {Phys. Rev. Lett.}\ }\textbf {\bibinfo {volume} {114}},\ \bibinfo {pages} {127401} (\bibinfo {year} {2015})}\BibitemShut {NoStop}%
\bibitem [{\citenamefont {Cheng}\ \emph {et~al.}(2016)\citenamefont {Cheng}, \citenamefont {Jouvaud}, \citenamefont {Ni}, \citenamefont {Mousavi}, \citenamefont {Genack},\ and\ \citenamefont {Khanikaev}}]{chengRobust2016}%
  \BibitemOpen
  \bibfield  {author} {\bibinfo {author} {\bibfnamefont {X.}~\bibnamefont {Cheng}}, \bibinfo {author} {\bibfnamefont {C.}~\bibnamefont {Jouvaud}}, \bibinfo {author} {\bibfnamefont {X.}~\bibnamefont {Ni}}, \bibinfo {author} {\bibfnamefont {S.~H.}\ \bibnamefont {Mousavi}}, \bibinfo {author} {\bibfnamefont {A.~Z.}\ \bibnamefont {Genack}},\ and\ \bibinfo {author} {\bibfnamefont {A.~B.}\ \bibnamefont {Khanikaev}},\ }\bibfield  {title} {\bibinfo {title} {Robust reconfigurable electromagnetic pathways within a photonic topological insulator},\ }\href {https://doi.org/10.1038/nmat4573} {\bibfield  {journal} {\bibinfo  {journal} {Nat. Mater}\ }\textbf {\bibinfo {volume} {15}},\ \bibinfo {pages} {542} (\bibinfo {year} {2016})}\BibitemShut {NoStop}%
\bibitem [{\citenamefont {Zhu}\ \emph {et~al.}(2018)\citenamefont {Zhu}, \citenamefont {Wang}, \citenamefont {Xu}, \citenamefont {Lai}, \citenamefont {Jiang},\ and\ \citenamefont {John}}]{zhuTopological2018}%
  \BibitemOpen
  \bibfield  {author} {\bibinfo {author} {\bibfnamefont {X.}~\bibnamefont {Zhu}}, \bibinfo {author} {\bibfnamefont {H.-X.}\ \bibnamefont {Wang}}, \bibinfo {author} {\bibfnamefont {C.}~\bibnamefont {Xu}}, \bibinfo {author} {\bibfnamefont {Y.}~\bibnamefont {Lai}}, \bibinfo {author} {\bibfnamefont {J.-H.}\ \bibnamefont {Jiang}},\ and\ \bibinfo {author} {\bibfnamefont {S.}~\bibnamefont {John}},\ }\bibfield  {title} {\bibinfo {title} {Topological transitions in continuously deformed photonic crystals},\ }\href {https://doi.org/10.1103/PhysRevB.97.085148} {\bibfield  {journal} {\bibinfo  {journal} {Phys. Rev. B}\ }\textbf {\bibinfo {volume} {97}},\ \bibinfo {pages} {085148} (\bibinfo {year} {2018})}\BibitemShut {NoStop}%
\bibitem [{\citenamefont {Barik}\ \emph {et~al.}(2016)\citenamefont {Barik}, \citenamefont {Miyake}, \citenamefont {DeGottardi}, \citenamefont {Waks},\ and\ \citenamefont {Hafezi}}]{barikTwodimensionally2016}%
  \BibitemOpen
  \bibfield  {author} {\bibinfo {author} {\bibfnamefont {S.}~\bibnamefont {Barik}}, \bibinfo {author} {\bibfnamefont {H.}~\bibnamefont {Miyake}}, \bibinfo {author} {\bibfnamefont {W.}~\bibnamefont {DeGottardi}}, \bibinfo {author} {\bibfnamefont {E.}~\bibnamefont {Waks}},\ and\ \bibinfo {author} {\bibfnamefont {M.}~\bibnamefont {Hafezi}},\ }\bibfield  {title} {\bibinfo {title} {Two-dimensionally confined topological edge states in photonic crystals},\ }\href {https://doi.org/10.1088/1367-2630/18/11/113013} {\bibfield  {journal} {\bibinfo  {journal} {New J. Phys.}\ }\textbf {\bibinfo {volume} {18}},\ \bibinfo {pages} {113013} (\bibinfo {year} {2016})}\BibitemShut {NoStop}%
\bibitem [{\citenamefont {Wu}\ and\ \citenamefont {Hu}(2015)}]{wuScheme2015}%
  \BibitemOpen
  \bibfield  {author} {\bibinfo {author} {\bibfnamefont {L.-H.}\ \bibnamefont {Wu}}\ and\ \bibinfo {author} {\bibfnamefont {X.}~\bibnamefont {Hu}},\ }\bibfield  {title} {\bibinfo {title} {Scheme for achieving a topological photonic crystal by using dielectric material},\ }\href@noop {} {\bibfield  {journal} {\bibinfo  {journal} {Physical review letters}\ }\textbf {\bibinfo {volume} {114}},\ \bibinfo {pages} {223901} (\bibinfo {year} {2015})}\BibitemShut {NoStop}%
\bibitem [{\citenamefont {Xu}\ \emph {et~al.}(2016)\citenamefont {Xu}, \citenamefont {Wang}, \citenamefont {Xu}, \citenamefont {Chen},\ and\ \citenamefont {Jiang}}]{xuAccidental2016}%
  \BibitemOpen
  \bibfield  {author} {\bibinfo {author} {\bibfnamefont {L.}~\bibnamefont {Xu}}, \bibinfo {author} {\bibfnamefont {H.-X.}\ \bibnamefont {Wang}}, \bibinfo {author} {\bibfnamefont {Y.-D.}\ \bibnamefont {Xu}}, \bibinfo {author} {\bibfnamefont {H.-Y.}\ \bibnamefont {Chen}},\ and\ \bibinfo {author} {\bibfnamefont {J.-H.}\ \bibnamefont {Jiang}},\ }\bibfield  {title} {\bibinfo {title} {Accidental degeneracy in photonic bands and topological phase transitions in two-dimensional core-shell dielectric photonic crystals},\ }\href {https://doi.org/10.1364/OE.24.018059} {\bibfield  {journal} {\bibinfo  {journal} {Opt. Express}\ }\textbf {\bibinfo {volume} {24}},\ \bibinfo {pages} {18059} (\bibinfo {year} {2016})}\BibitemShut {NoStop}%
\bibitem [{\citenamefont {Wang}\ \emph {et~al.}(2016)\citenamefont {Wang}, \citenamefont {Xu}, \citenamefont {Chen},\ and\ \citenamefont {Jiang}}]{wangThreedimensional2016}%
  \BibitemOpen
  \bibfield  {author} {\bibinfo {author} {\bibfnamefont {H.}~\bibnamefont {Wang}}, \bibinfo {author} {\bibfnamefont {L.}~\bibnamefont {Xu}}, \bibinfo {author} {\bibfnamefont {H.}~\bibnamefont {Chen}},\ and\ \bibinfo {author} {\bibfnamefont {J.-H.}\ \bibnamefont {Jiang}},\ }\bibfield  {title} {\bibinfo {title} {Three-dimensional photonic {{Dirac}} points stabilized by point group symmetry},\ }\href {https://doi.org/10.1103/PhysRevB.93.235155} {\bibfield  {journal} {\bibinfo  {journal} {Phys. Rev. B}\ }\textbf {\bibinfo {volume} {93}},\ \bibinfo {pages} {235155} (\bibinfo {year} {2016})}\BibitemShut {NoStop}%
\bibitem [{\citenamefont {Lu}\ \emph {et~al.}(2016)\citenamefont {Lu}, \citenamefont {Fang}, \citenamefont {Fu}, \citenamefont {Johnson}, \citenamefont {Joannopoulos},\ and\ \citenamefont {Solja{\v c}i{\'c}}}]{luSymmetryprotected2016}%
  \BibitemOpen
  \bibfield  {author} {\bibinfo {author} {\bibfnamefont {L.}~\bibnamefont {Lu}}, \bibinfo {author} {\bibfnamefont {C.}~\bibnamefont {Fang}}, \bibinfo {author} {\bibfnamefont {L.}~\bibnamefont {Fu}}, \bibinfo {author} {\bibfnamefont {S.~G.}\ \bibnamefont {Johnson}}, \bibinfo {author} {\bibfnamefont {J.~D.}\ \bibnamefont {Joannopoulos}},\ and\ \bibinfo {author} {\bibfnamefont {M.}~\bibnamefont {Solja{\v c}i{\'c}}},\ }\bibfield  {title} {\bibinfo {title} {Symmetry-protected topological photonic crystal in three dimensions},\ }\href {https://doi.org/10.1038/nphys3611} {\bibfield  {journal} {\bibinfo  {journal} {Nat. Phys.}\ }\textbf {\bibinfo {volume} {12}},\ \bibinfo {pages} {337} (\bibinfo {year} {2016})}\BibitemShut {NoStop}%
\bibitem [{\citenamefont {Gao}\ \emph {et~al.}(2018)\citenamefont {Gao}, \citenamefont {Xue}, \citenamefont {Yang}, \citenamefont {Lai}, \citenamefont {Yu}, \citenamefont {Lin}, \citenamefont {Chong}, \citenamefont {Shvets},\ and\ \citenamefont {Zhang}}]{gaoTopologically2018}%
  \BibitemOpen
  \bibfield  {author} {\bibinfo {author} {\bibfnamefont {F.}~\bibnamefont {Gao}}, \bibinfo {author} {\bibfnamefont {H.}~\bibnamefont {Xue}}, \bibinfo {author} {\bibfnamefont {Z.}~\bibnamefont {Yang}}, \bibinfo {author} {\bibfnamefont {K.}~\bibnamefont {Lai}}, \bibinfo {author} {\bibfnamefont {Y.}~\bibnamefont {Yu}}, \bibinfo {author} {\bibfnamefont {X.}~\bibnamefont {Lin}}, \bibinfo {author} {\bibfnamefont {Y.}~\bibnamefont {Chong}}, \bibinfo {author} {\bibfnamefont {G.}~\bibnamefont {Shvets}},\ and\ \bibinfo {author} {\bibfnamefont {B.}~\bibnamefont {Zhang}},\ }\bibfield  {title} {\bibinfo {title} {Topologically protected refraction of robust kink states in valley photonic crystals},\ }\href {https://doi.org/10.1038/nphys4304} {\bibfield  {journal} {\bibinfo  {journal} {Nat. Phys.}\ }\textbf {\bibinfo {volume} {14}},\ \bibinfo {pages} {140} (\bibinfo {year} {2018})}\BibitemShut {NoStop}%
\bibitem [{\citenamefont {Shalaev}\ \emph {et~al.}(2019)\citenamefont {Shalaev}, \citenamefont {Walasik}, \citenamefont {Tsukernik}, \citenamefont {Xu},\ and\ \citenamefont {Litchinitser}}]{shalaevRobust2019}%
  \BibitemOpen
  \bibfield  {author} {\bibinfo {author} {\bibfnamefont {M.~I.}\ \bibnamefont {Shalaev}}, \bibinfo {author} {\bibfnamefont {W.}~\bibnamefont {Walasik}}, \bibinfo {author} {\bibfnamefont {A.}~\bibnamefont {Tsukernik}}, \bibinfo {author} {\bibfnamefont {Y.}~\bibnamefont {Xu}},\ and\ \bibinfo {author} {\bibfnamefont {N.~M.}\ \bibnamefont {Litchinitser}},\ }\bibfield  {title} {\bibinfo {title} {Robust topologically protected transport in photonic crystals at telecommunication wavelengths},\ }\href {https://doi.org/10.1038/s41565-018-0297-6} {\bibfield  {journal} {\bibinfo  {journal} {Nat. Nanotechnol}\ }\textbf {\bibinfo {volume} {14}},\ \bibinfo {pages} {31} (\bibinfo {year} {2019})}\BibitemShut {NoStop}%
\bibitem [{\citenamefont {Ryu}\ and\ \citenamefont {Hatsugai}(2002)}]{ryuTopological2002}%
  \BibitemOpen
  \bibfield  {author} {\bibinfo {author} {\bibfnamefont {S.}~\bibnamefont {Ryu}}\ and\ \bibinfo {author} {\bibfnamefont {Y.}~\bibnamefont {Hatsugai}},\ }\bibfield  {title} {\bibinfo {title} {Topological {{Origin}} of {{Zero-Energy Edge States}} in {{Particle-Hole Symmetric Systems}}},\ }\href {https://doi.org/10.1103/PhysRevLett.89.077002} {\bibfield  {journal} {\bibinfo  {journal} {Phys. Rev. Lett.}\ }\textbf {\bibinfo {volume} {89}},\ \bibinfo {pages} {077002} (\bibinfo {year} {2002})}\BibitemShut {NoStop}%
\bibitem [{\citenamefont {Xie}\ \emph {et~al.}(2018)\citenamefont {Xie}, \citenamefont {Wang}, \citenamefont {Wang}, \citenamefont {Zhu}, \citenamefont {Jiang}, \citenamefont {Lu},\ and\ \citenamefont {Chen}}]{xieSecondorder2018}%
  \BibitemOpen
  \bibfield  {author} {\bibinfo {author} {\bibfnamefont {B.-Y.}\ \bibnamefont {Xie}}, \bibinfo {author} {\bibfnamefont {H.-F.}\ \bibnamefont {Wang}}, \bibinfo {author} {\bibfnamefont {H.-X.}\ \bibnamefont {Wang}}, \bibinfo {author} {\bibfnamefont {X.-Y.}\ \bibnamefont {Zhu}}, \bibinfo {author} {\bibfnamefont {J.-H.}\ \bibnamefont {Jiang}}, \bibinfo {author} {\bibfnamefont {M.-H.}\ \bibnamefont {Lu}},\ and\ \bibinfo {author} {\bibfnamefont {Y.-F.}\ \bibnamefont {Chen}},\ }\bibfield  {title} {\bibinfo {title} {Second-order photonic topological insulator with corner states},\ }\href {https://doi.org/10.1103/PhysRevB.98.205147} {\bibfield  {journal} {\bibinfo  {journal} {Phys. Rev. B}\ }\textbf {\bibinfo {volume} {98}},\ \bibinfo {pages} {205147} (\bibinfo {year} {2018})}\BibitemShut {NoStop}%
\bibitem [{\citenamefont {Xie}\ \emph {et~al.}(2020)\citenamefont {Xie}, \citenamefont {Su}, \citenamefont {Wang}, \citenamefont {Liu}, \citenamefont {Hu}, \citenamefont {Yu}, \citenamefont {Zhan}, \citenamefont {Lu}, \citenamefont {Wang},\ and\ \citenamefont {Chen}}]{xieHigherorder2020}%
  \BibitemOpen
  \bibfield  {author} {\bibinfo {author} {\bibfnamefont {B.}~\bibnamefont {Xie}}, \bibinfo {author} {\bibfnamefont {G.}~\bibnamefont {Su}}, \bibinfo {author} {\bibfnamefont {H.-F.}\ \bibnamefont {Wang}}, \bibinfo {author} {\bibfnamefont {F.}~\bibnamefont {Liu}}, \bibinfo {author} {\bibfnamefont {L.}~\bibnamefont {Hu}}, \bibinfo {author} {\bibfnamefont {S.-Y.}\ \bibnamefont {Yu}}, \bibinfo {author} {\bibfnamefont {P.}~\bibnamefont {Zhan}}, \bibinfo {author} {\bibfnamefont {M.-H.}\ \bibnamefont {Lu}}, \bibinfo {author} {\bibfnamefont {Z.}~\bibnamefont {Wang}},\ and\ \bibinfo {author} {\bibfnamefont {Y.-F.}\ \bibnamefont {Chen}},\ }\bibfield  {title} {\bibinfo {title} {Higher-order quantum spin {{Hall}} effect in a photonic crystal},\ }\href {https://doi.org/10.1038/s41467-020-17593-8} {\bibfield  {journal} {\bibinfo  {journal} {Nat. Commun}\ }\textbf {\bibinfo {volume} {11}},\ \bibinfo {pages} {3768} (\bibinfo {year} {2020})}\BibitemShut {NoStop}%
\bibitem [{\citenamefont {Xie}\ \emph {et~al.}(2019)\citenamefont {Xie}, \citenamefont {Su}, \citenamefont {Wang}, \citenamefont {Su}, \citenamefont {Shen}, \citenamefont {Zhan}, \citenamefont {Lu}, \citenamefont {Wang},\ and\ \citenamefont {Chen}}]{xieVisualization2019}%
  \BibitemOpen
  \bibfield  {author} {\bibinfo {author} {\bibfnamefont {B.-Y.}\ \bibnamefont {Xie}}, \bibinfo {author} {\bibfnamefont {G.-X.}\ \bibnamefont {Su}}, \bibinfo {author} {\bibfnamefont {H.-F.}\ \bibnamefont {Wang}}, \bibinfo {author} {\bibfnamefont {H.}~\bibnamefont {Su}}, \bibinfo {author} {\bibfnamefont {X.-P.}\ \bibnamefont {Shen}}, \bibinfo {author} {\bibfnamefont {P.}~\bibnamefont {Zhan}}, \bibinfo {author} {\bibfnamefont {M.-H.}\ \bibnamefont {Lu}}, \bibinfo {author} {\bibfnamefont {Z.-L.}\ \bibnamefont {Wang}},\ and\ \bibinfo {author} {\bibfnamefont {Y.-F.}\ \bibnamefont {Chen}},\ }\bibfield  {title} {\bibinfo {title} {Visualization of {{Higher-Order Topological Insulating Phases}} in {{Two-Dimensional Dielectric Photonic Crystals}}},\ }\href {https://doi.org/10.1103/PhysRevLett.122.233903} {\bibfield  {journal} {\bibinfo  {journal} {Phys. Rev. Lett.}\ }\textbf {\bibinfo {volume} {122}},\ \bibinfo {pages} {233903} (\bibinfo {year} {2019})}\BibitemShut {NoStop}%
\bibitem [{\citenamefont {Xie}\ \emph {et~al.}(2021)\citenamefont {Xie}, \citenamefont {Wang}, \citenamefont {Zhang}, \citenamefont {Zhan}, \citenamefont {Jiang}, \citenamefont {Lu},\ and\ \citenamefont {Chen}}]{xieHigherorder2021}%
  \BibitemOpen
  \bibfield  {author} {\bibinfo {author} {\bibfnamefont {B.}~\bibnamefont {Xie}}, \bibinfo {author} {\bibfnamefont {H.-X.}\ \bibnamefont {Wang}}, \bibinfo {author} {\bibfnamefont {X.}~\bibnamefont {Zhang}}, \bibinfo {author} {\bibfnamefont {P.}~\bibnamefont {Zhan}}, \bibinfo {author} {\bibfnamefont {J.-H.}\ \bibnamefont {Jiang}}, \bibinfo {author} {\bibfnamefont {M.}~\bibnamefont {Lu}},\ and\ \bibinfo {author} {\bibfnamefont {Y.}~\bibnamefont {Chen}},\ }\bibfield  {title} {\bibinfo {title} {Higher-order band topology},\ }\href {https://doi.org/10.1038/s42254-021-00323-4} {\bibfield  {journal} {\bibinfo  {journal} {Nat. Rev. Phys}\ }\textbf {\bibinfo {volume} {3}},\ \bibinfo {pages} {520} (\bibinfo {year} {2021})}\BibitemShut {NoStop}%
\bibitem [{\citenamefont {Benalcazar}\ \emph {et~al.}(2017{\natexlab{a}})\citenamefont {Benalcazar}, \citenamefont {Bernevig},\ and\ \citenamefont {Hughes}}]{benalcazarQuantized2017}%
  \BibitemOpen
  \bibfield  {author} {\bibinfo {author} {\bibfnamefont {W.~A.}\ \bibnamefont {Benalcazar}}, \bibinfo {author} {\bibfnamefont {B.~A.}\ \bibnamefont {Bernevig}},\ and\ \bibinfo {author} {\bibfnamefont {T.~L.}\ \bibnamefont {Hughes}},\ }\bibfield  {title} {\bibinfo {title} {Quantized electric multipole insulators},\ }\href {https://doi.org/10.1126/science.aah6442} {\bibfield  {journal} {\bibinfo  {journal} {Science}\ }\textbf {\bibinfo {volume} {357}},\ \bibinfo {pages} {61} (\bibinfo {year} {2017}{\natexlab{a}})}\BibitemShut {NoStop}%
\bibitem [{\citenamefont {Langbehn}\ \emph {et~al.}(2017)\citenamefont {Langbehn}, \citenamefont {Peng}, \citenamefont {Trifunovic}, \citenamefont {Von~Oppen},\ and\ \citenamefont {Brouwer}}]{langbehnReflectionSymmetric2017}%
  \BibitemOpen
  \bibfield  {author} {\bibinfo {author} {\bibfnamefont {J.}~\bibnamefont {Langbehn}}, \bibinfo {author} {\bibfnamefont {Y.}~\bibnamefont {Peng}}, \bibinfo {author} {\bibfnamefont {L.}~\bibnamefont {Trifunovic}}, \bibinfo {author} {\bibfnamefont {F.}~\bibnamefont {Von~Oppen}},\ and\ \bibinfo {author} {\bibfnamefont {P.~W.}\ \bibnamefont {Brouwer}},\ }\bibfield  {title} {\bibinfo {title} {Reflection-{{Symmetric Second-Order Topological Insulators}} and {{Superconductors}}},\ }\href {https://doi.org/10.1103/PhysRevLett.119.246401} {\bibfield  {journal} {\bibinfo  {journal} {Phys. Rev. Lett.}\ }\textbf {\bibinfo {volume} {119}},\ \bibinfo {pages} {246401} (\bibinfo {year} {2017})}\BibitemShut {NoStop}%
\bibitem [{\citenamefont {Song}\ \emph {et~al.}(2017)\citenamefont {Song}, \citenamefont {Fang},\ and\ \citenamefont {Fang}}]{songDimensional2017}%
  \BibitemOpen
  \bibfield  {author} {\bibinfo {author} {\bibfnamefont {Z.}~\bibnamefont {Song}}, \bibinfo {author} {\bibfnamefont {Z.}~\bibnamefont {Fang}},\ and\ \bibinfo {author} {\bibfnamefont {C.}~\bibnamefont {Fang}},\ }\bibfield  {title} {\bibinfo {title} {( d - 2 ) -{{Dimensional Edge States}} of {{Rotation Symmetry Protected Topological States}}},\ }\href {https://doi.org/10.1103/PhysRevLett.119.246402} {\bibfield  {journal} {\bibinfo  {journal} {Phys. Rev. Lett.}\ }\textbf {\bibinfo {volume} {119}},\ \bibinfo {pages} {246402} (\bibinfo {year} {2017})}\BibitemShut {NoStop}%
\bibitem [{\citenamefont {Benalcazar}\ \emph {et~al.}(2017{\natexlab{b}})\citenamefont {Benalcazar}, \citenamefont {Bernevig},\ and\ \citenamefont {Hughes}}]{benalcazarElectric2017}%
  \BibitemOpen
  \bibfield  {author} {\bibinfo {author} {\bibfnamefont {W.~A.}\ \bibnamefont {Benalcazar}}, \bibinfo {author} {\bibfnamefont {B.~A.}\ \bibnamefont {Bernevig}},\ and\ \bibinfo {author} {\bibfnamefont {T.~L.}\ \bibnamefont {Hughes}},\ }\bibfield  {title} {\bibinfo {title} {Electric multipole moments, topological multipole moment pumping, and chiral hinge states in crystalline insulators},\ }\href {https://doi.org/10.1103/PhysRevB.96.245115} {\bibfield  {journal} {\bibinfo  {journal} {Phys. Rev. B}\ }\textbf {\bibinfo {volume} {96}},\ \bibinfo {pages} {245115} (\bibinfo {year} {2017}{\natexlab{b}})}\BibitemShut {NoStop}%
\bibitem [{\citenamefont {Zhang}\ \emph {et~al.}(2019)\citenamefont {Zhang}, \citenamefont {Jiang}, \citenamefont {Song}, \citenamefont {Huang}, \citenamefont {He}, \citenamefont {Fang}, \citenamefont {Weng},\ and\ \citenamefont {Fang}}]{zhangCatalogue2019}%
  \BibitemOpen
  \bibfield  {author} {\bibinfo {author} {\bibfnamefont {T.}~\bibnamefont {Zhang}}, \bibinfo {author} {\bibfnamefont {Y.}~\bibnamefont {Jiang}}, \bibinfo {author} {\bibfnamefont {Z.}~\bibnamefont {Song}}, \bibinfo {author} {\bibfnamefont {H.}~\bibnamefont {Huang}}, \bibinfo {author} {\bibfnamefont {Y.}~\bibnamefont {He}}, \bibinfo {author} {\bibfnamefont {Z.}~\bibnamefont {Fang}}, \bibinfo {author} {\bibfnamefont {H.}~\bibnamefont {Weng}},\ and\ \bibinfo {author} {\bibfnamefont {C.}~\bibnamefont {Fang}},\ }\bibfield  {title} {\bibinfo {title} {Catalogue of topological electronic materials},\ }\href {https://doi.org/10.1038/s41586-019-0944-6} {\bibfield  {journal} {\bibinfo  {journal} {Nature}\ }\textbf {\bibinfo {volume} {566}},\ \bibinfo {pages} {475} (\bibinfo {year} {2019})}\BibitemShut {NoStop}%
\bibitem [{\citenamefont {Fang}\ \emph {et~al.}(2003)\citenamefont {Fang}, \citenamefont {Nagaosa}, \citenamefont {Takahashi}, \citenamefont {Asamitsu}, \citenamefont {Mathieu}, \citenamefont {Ogasawara}, \citenamefont {Yamada}, \citenamefont {Kawasaki}, \citenamefont {Tokura},\ and\ \citenamefont {Terakura}}]{fangAnomalous2003}%
  \BibitemOpen
  \bibfield  {author} {\bibinfo {author} {\bibfnamefont {Z.}~\bibnamefont {Fang}}, \bibinfo {author} {\bibfnamefont {N.}~\bibnamefont {Nagaosa}}, \bibinfo {author} {\bibfnamefont {K.~S.}\ \bibnamefont {Takahashi}}, \bibinfo {author} {\bibfnamefont {A.}~\bibnamefont {Asamitsu}}, \bibinfo {author} {\bibfnamefont {R.}~\bibnamefont {Mathieu}}, \bibinfo {author} {\bibfnamefont {T.}~\bibnamefont {Ogasawara}}, \bibinfo {author} {\bibfnamefont {H.}~\bibnamefont {Yamada}}, \bibinfo {author} {\bibfnamefont {M.}~\bibnamefont {Kawasaki}}, \bibinfo {author} {\bibfnamefont {Y.}~\bibnamefont {Tokura}},\ and\ \bibinfo {author} {\bibfnamefont {K.}~\bibnamefont {Terakura}},\ }\bibfield  {title} {\bibinfo {title} {The {{Anomalous Hall Effect}} and {{Magnetic Monopoles}} in {{Momentum Space}}},\ }\href {https://doi.org/10.1126/science.1089408} {\bibfield  {journal} {\bibinfo  {journal} {Science}\ }\textbf {\bibinfo {volume} {302}},\ \bibinfo {pages} {92} (\bibinfo {year} {2003})}\BibitemShut {NoStop}%
\bibitem [{\citenamefont {Yang}\ \emph {et~al.}(2020)\citenamefont {Yang}, \citenamefont {Yamagami}, \citenamefont {Yu}, \citenamefont {Pitchappa}, \citenamefont {Webber}, \citenamefont {Zhang}, \citenamefont {Fujita}, \citenamefont {Nagatsuma},\ and\ \citenamefont {Singh}}]{yangTerahertz2020}%
  \BibitemOpen
  \bibfield  {author} {\bibinfo {author} {\bibfnamefont {Y.}~\bibnamefont {Yang}}, \bibinfo {author} {\bibfnamefont {Y.}~\bibnamefont {Yamagami}}, \bibinfo {author} {\bibfnamefont {X.}~\bibnamefont {Yu}}, \bibinfo {author} {\bibfnamefont {P.}~\bibnamefont {Pitchappa}}, \bibinfo {author} {\bibfnamefont {J.}~\bibnamefont {Webber}}, \bibinfo {author} {\bibfnamefont {B.}~\bibnamefont {Zhang}}, \bibinfo {author} {\bibfnamefont {M.}~\bibnamefont {Fujita}}, \bibinfo {author} {\bibfnamefont {T.}~\bibnamefont {Nagatsuma}},\ and\ \bibinfo {author} {\bibfnamefont {R.}~\bibnamefont {Singh}},\ }\bibfield  {title} {\bibinfo {title} {Terahertz topological photonics for on-chip communication},\ }\href {https://doi.org/10.1038/s41566-020-0618-9} {\bibfield  {journal} {\bibinfo  {journal} {Nat. Photonics}\ }\textbf {\bibinfo {volume} {14}},\ \bibinfo {pages} {446} (\bibinfo {year} {2020})}\BibitemShut {NoStop}%
\bibitem [{\citenamefont {Zeng}\ \emph {et~al.}(2020)\citenamefont {Zeng}, \citenamefont {Chattopadhyay}, \citenamefont {Zhu}, \citenamefont {Qiang}, \citenamefont {Li}, \citenamefont {Jin}, \citenamefont {Li}, \citenamefont {Davies}, \citenamefont {Linfield}, \citenamefont {Zhang}, \citenamefont {Chong},\ and\ \citenamefont {Wang}}]{zengElectrically2020}%
  \BibitemOpen
  \bibfield  {author} {\bibinfo {author} {\bibfnamefont {Y.}~\bibnamefont {Zeng}}, \bibinfo {author} {\bibfnamefont {U.}~\bibnamefont {Chattopadhyay}}, \bibinfo {author} {\bibfnamefont {B.}~\bibnamefont {Zhu}}, \bibinfo {author} {\bibfnamefont {B.}~\bibnamefont {Qiang}}, \bibinfo {author} {\bibfnamefont {J.}~\bibnamefont {Li}}, \bibinfo {author} {\bibfnamefont {Y.}~\bibnamefont {Jin}}, \bibinfo {author} {\bibfnamefont {L.}~\bibnamefont {Li}}, \bibinfo {author} {\bibfnamefont {A.~G.}\ \bibnamefont {Davies}}, \bibinfo {author} {\bibfnamefont {E.~H.}\ \bibnamefont {Linfield}}, \bibinfo {author} {\bibfnamefont {B.}~\bibnamefont {Zhang}}, \bibinfo {author} {\bibfnamefont {Y.}~\bibnamefont {Chong}},\ and\ \bibinfo {author} {\bibfnamefont {Q.~J.}\ \bibnamefont {Wang}},\ }\bibfield  {title} {\bibinfo {title} {Electrically pumped topological laser with valley edge modes},\ }\href {https://doi.org/10.1038/s41586-020-1981-x} {\bibfield  {journal} {\bibinfo  {journal} {Nature}\ }\textbf {\bibinfo {volume} {578}},\
  \bibinfo {pages} {246} (\bibinfo {year} {2020})}\BibitemShut {NoStop}%
\bibitem [{\citenamefont {Ota}\ \emph {et~al.}(2018)\citenamefont {Ota}, \citenamefont {Katsumi}, \citenamefont {Watanabe}, \citenamefont {Iwamoto},\ and\ \citenamefont {Arakawa}}]{Ota.K.W.I.A2018}%
  \BibitemOpen
  \bibfield  {author} {\bibinfo {author} {\bibfnamefont {Y.}~\bibnamefont {Ota}}, \bibinfo {author} {\bibfnamefont {R.}~\bibnamefont {Katsumi}}, \bibinfo {author} {\bibfnamefont {K.}~\bibnamefont {Watanabe}}, \bibinfo {author} {\bibfnamefont {S.}~\bibnamefont {Iwamoto}},\ and\ \bibinfo {author} {\bibfnamefont {Y.}~\bibnamefont {Arakawa}},\ }\bibfield  {title} {\bibinfo {title} {Topological photonic crystal nanocavity laser},\ }\href {https://doi.org/10.1038/s42005-018-0083-7} {\bibfield  {journal} {\bibinfo  {journal} {Commun. Phys.}\ }\textbf {\bibinfo {volume} {1}},\ \bibinfo {pages} {86} (\bibinfo {year} {2018})}\BibitemShut {NoStop}%
\bibitem [{\citenamefont {Bandres}\ \emph {et~al.}(2018)\citenamefont {Bandres}, \citenamefont {Wittek}, \citenamefont {Harari}, \citenamefont {Parto}, \citenamefont {Ren}, \citenamefont {Segev}, \citenamefont {Christodoulides},\ and\ \citenamefont {Khajavikhan}}]{bandresTopological2018}%
  \BibitemOpen
  \bibfield  {author} {\bibinfo {author} {\bibfnamefont {M.~A.}\ \bibnamefont {Bandres}}, \bibinfo {author} {\bibfnamefont {S.}~\bibnamefont {Wittek}}, \bibinfo {author} {\bibfnamefont {G.}~\bibnamefont {Harari}}, \bibinfo {author} {\bibfnamefont {M.}~\bibnamefont {Parto}}, \bibinfo {author} {\bibfnamefont {J.}~\bibnamefont {Ren}}, \bibinfo {author} {\bibfnamefont {M.}~\bibnamefont {Segev}}, \bibinfo {author} {\bibfnamefont {D.~N.}\ \bibnamefont {Christodoulides}},\ and\ \bibinfo {author} {\bibfnamefont {M.}~\bibnamefont {Khajavikhan}},\ }\bibfield  {title} {\bibinfo {title} {Topological insulator laser: {{Experiments}}},\ }\href {https://doi.org/10.1126/science.aar4005} {\bibfield  {journal} {\bibinfo  {journal} {Science}\ }\textbf {\bibinfo {volume} {359}},\ \bibinfo {pages} {eaar4005} (\bibinfo {year} {2018})}\BibitemShut {NoStop}%
\bibitem [{\citenamefont {Miyake}(2010)}]{miyakeQuantum2010}%
  \BibitemOpen
  \bibfield  {author} {\bibinfo {author} {\bibfnamefont {A.}~\bibnamefont {Miyake}},\ }\bibfield  {title} {\bibinfo {title} {Quantum {{Computation}} on the {{Edge}} of a {{Symmetry-Protected Topological Order}}},\ }\href {https://doi.org/10.1103/PhysRevLett.105.040501} {\bibfield  {journal} {\bibinfo  {journal} {Phys. Rev. Lett.}\ }\textbf {\bibinfo {volume} {105}},\ \bibinfo {pages} {040501} (\bibinfo {year} {2010})}\BibitemShut {NoStop}%
\bibitem [{\citenamefont {Else}\ \emph {et~al.}(2012)\citenamefont {Else}, \citenamefont {Schwarz}, \citenamefont {Bartlett},\ and\ \citenamefont {Doherty}}]{elseSymmetryProtected2012}%
  \BibitemOpen
  \bibfield  {author} {\bibinfo {author} {\bibfnamefont {D.~V.}\ \bibnamefont {Else}}, \bibinfo {author} {\bibfnamefont {I.}~\bibnamefont {Schwarz}}, \bibinfo {author} {\bibfnamefont {S.~D.}\ \bibnamefont {Bartlett}},\ and\ \bibinfo {author} {\bibfnamefont {A.~C.}\ \bibnamefont {Doherty}},\ }\bibfield  {title} {\bibinfo {title} {Symmetry-{{Protected Phases}} for {{Measurement-Based Quantum Computation}}},\ }\href {https://doi.org/10.1103/PhysRevLett.108.240505} {\bibfield  {journal} {\bibinfo  {journal} {Phys. Rev. Lett.}\ }\textbf {\bibinfo {volume} {108}},\ \bibinfo {pages} {240505} (\bibinfo {year} {2012})}\BibitemShut {NoStop}%
\bibitem [{\citenamefont {Wang}\ \emph {et~al.}(2020)\citenamefont {Wang}, \citenamefont {Zhou}, \citenamefont {Bi}, \citenamefont {Qiu}, \citenamefont {Ke},\ and\ \citenamefont {Liu}}]{wangValleylocked2020}%
  \BibitemOpen
  \bibfield  {author} {\bibinfo {author} {\bibfnamefont {M.}~\bibnamefont {Wang}}, \bibinfo {author} {\bibfnamefont {W.}~\bibnamefont {Zhou}}, \bibinfo {author} {\bibfnamefont {L.}~\bibnamefont {Bi}}, \bibinfo {author} {\bibfnamefont {C.}~\bibnamefont {Qiu}}, \bibinfo {author} {\bibfnamefont {M.}~\bibnamefont {Ke}},\ and\ \bibinfo {author} {\bibfnamefont {Z.}~\bibnamefont {Liu}},\ }\bibfield  {title} {\bibinfo {title} {Valley-locked waveguide transport in acoustic heterostructures},\ }\href {https://doi.org/10.1038/s41467-020-16843-z} {\bibfield  {journal} {\bibinfo  {journal} {Nat. Commun}\ }\textbf {\bibinfo {volume} {11}},\ \bibinfo {pages} {3000} (\bibinfo {year} {2020})}\BibitemShut {NoStop}%
\bibitem [{\citenamefont {Wang}\ \emph {et~al.}(2021)\citenamefont {Wang}, \citenamefont {Zhang}, \citenamefont {Zhang}, \citenamefont {Wang}, \citenamefont {Guo}, \citenamefont {Zhang},\ and\ \citenamefont {Chan}}]{wangTopological2021a}%
  \BibitemOpen
  \bibfield  {author} {\bibinfo {author} {\bibfnamefont {M.}~\bibnamefont {Wang}}, \bibinfo {author} {\bibfnamefont {R.-Y.}\ \bibnamefont {Zhang}}, \bibinfo {author} {\bibfnamefont {L.}~\bibnamefont {Zhang}}, \bibinfo {author} {\bibfnamefont {D.}~\bibnamefont {Wang}}, \bibinfo {author} {\bibfnamefont {Q.}~\bibnamefont {Guo}}, \bibinfo {author} {\bibfnamefont {Z.-Q.}\ \bibnamefont {Zhang}},\ and\ \bibinfo {author} {\bibfnamefont {C.~T.}\ \bibnamefont {Chan}},\ }\bibfield  {title} {\bibinfo {title} {Topological {{One-Way Large-Area Waveguide States}} in {{Magnetic Photonic Crystals}}},\ }\href {https://doi.org/10.1103/PhysRevLett.126.067401} {\bibfield  {journal} {\bibinfo  {journal} {Phys. Rev. Lett.}\ }\textbf {\bibinfo {volume} {126}},\ \bibinfo {pages} {067401} (\bibinfo {year} {2021})}\BibitemShut {NoStop}%
\bibitem [{\citenamefont {Chen}\ \emph {et~al.}(2021)\citenamefont {Chen}, \citenamefont {Zhang}, \citenamefont {Chen}, \citenamefont {Yan}, \citenamefont {Xi}, \citenamefont {Chen},\ and\ \citenamefont {Yang}}]{chenPhotonic2021}%
  \BibitemOpen
  \bibfield  {author} {\bibinfo {author} {\bibfnamefont {Q.}~\bibnamefont {Chen}}, \bibinfo {author} {\bibfnamefont {L.}~\bibnamefont {Zhang}}, \bibinfo {author} {\bibfnamefont {F.}~\bibnamefont {Chen}}, \bibinfo {author} {\bibfnamefont {Q.}~\bibnamefont {Yan}}, \bibinfo {author} {\bibfnamefont {R.}~\bibnamefont {Xi}}, \bibinfo {author} {\bibfnamefont {H.}~\bibnamefont {Chen}},\ and\ \bibinfo {author} {\bibfnamefont {Y.}~\bibnamefont {Yang}},\ }\bibfield  {title} {\bibinfo {title} {Photonic {{Topological Valley-Locked Waveguides}}},\ }\href {https://doi.org/10.1021/acsphotonics.1c00029} {\bibfield  {journal} {\bibinfo  {journal} {ACS Photonics}\ }\textbf {\bibinfo {volume} {8}},\ \bibinfo {pages} {1400} (\bibinfo {year} {2021})}\BibitemShut {NoStop}%
\bibitem [{\citenamefont {Wang}\ \emph {et~al.}(2024)\citenamefont {Wang}, \citenamefont {Zhang}, \citenamefont {Zhang}, \citenamefont {Xue}, \citenamefont {Jia}, \citenamefont {Hu}, \citenamefont {Wang}, \citenamefont {Jiang},\ and\ \citenamefont {Chan}}]{wang2024TI}%
  \BibitemOpen
  \bibfield  {author} {\bibinfo {author} {\bibfnamefont {M.}~\bibnamefont {Wang}}, \bibinfo {author} {\bibfnamefont {R.-Y.}\ \bibnamefont {Zhang}}, \bibinfo {author} {\bibfnamefont {C.}~\bibnamefont {Zhang}}, \bibinfo {author} {\bibfnamefont {H.}~\bibnamefont {Xue}}, \bibinfo {author} {\bibfnamefont {H.}~\bibnamefont {Jia}}, \bibinfo {author} {\bibfnamefont {J.}~\bibnamefont {Hu}}, \bibinfo {author} {\bibfnamefont {D.}~\bibnamefont {Wang}}, \bibinfo {author} {\bibfnamefont {T.}~\bibnamefont {Jiang}},\ and\ \bibinfo {author} {\bibfnamefont {C.~T.}\ \bibnamefont {Chan}},\ }\href {https://arxiv.org/abs/2407.00440} {\bibinfo {title} {Three-dimensional non-reciprocal transport in photonic topological heterostructure of arbitrary shape}} (\bibinfo {year} {2024}),\ \Eprint {https://arxiv.org/abs/2407.00440} {arXiv:2407.00440 [physics.optics]} \BibitemShut {NoStop}%
\bibitem [{\citenamefont {Zhao}\ \emph {et~al.}(2019)\citenamefont {Zhao}, \citenamefont {Qiao}, \citenamefont {Wu}, \citenamefont {Midya}, \citenamefont {Longhi},\ and\ \citenamefont {Feng}}]{zhao_non-hermitian_2019}%
  \BibitemOpen
  \bibfield  {author} {\bibinfo {author} {\bibfnamefont {H.}~\bibnamefont {Zhao}}, \bibinfo {author} {\bibfnamefont {X.}~\bibnamefont {Qiao}}, \bibinfo {author} {\bibfnamefont {T.}~\bibnamefont {Wu}}, \bibinfo {author} {\bibfnamefont {B.}~\bibnamefont {Midya}}, \bibinfo {author} {\bibfnamefont {S.}~\bibnamefont {Longhi}},\ and\ \bibinfo {author} {\bibfnamefont {L.}~\bibnamefont {Feng}},\ }\bibfield  {title} {\bibinfo {title} {Non-hermitian topological light steering},\ }\href {https://doi.org/10.1126/science.aay1064} {\bibfield  {journal} {\bibinfo  {journal} {Science}\ }\textbf {\bibinfo {volume} {365}},\ \bibinfo {pages} {1163} (\bibinfo {year} {2019})},\ \bibinfo {note} {publisher: American Association for the Advancement of Science}\BibitemShut {NoStop}%
\bibitem [{\citenamefont {Bahari}\ \emph {et~al.}(2021)\citenamefont {Bahari}, \citenamefont {Hsu}, \citenamefont {Pan}, \citenamefont {Preece}, \citenamefont {Ndao}, \citenamefont {El~Amili}, \citenamefont {Fainman},\ and\ \citenamefont {Kanté}}]{bahari_photonic_2021}%
  \BibitemOpen
  \bibfield  {author} {\bibinfo {author} {\bibfnamefont {B.}~\bibnamefont {Bahari}}, \bibinfo {author} {\bibfnamefont {L.}~\bibnamefont {Hsu}}, \bibinfo {author} {\bibfnamefont {S.~H.}\ \bibnamefont {Pan}}, \bibinfo {author} {\bibfnamefont {D.}~\bibnamefont {Preece}}, \bibinfo {author} {\bibfnamefont {A.}~\bibnamefont {Ndao}}, \bibinfo {author} {\bibfnamefont {A.}~\bibnamefont {El~Amili}}, \bibinfo {author} {\bibfnamefont {Y.}~\bibnamefont {Fainman}},\ and\ \bibinfo {author} {\bibfnamefont {B.}~\bibnamefont {Kanté}},\ }\bibfield  {title} {\bibinfo {title} {Photonic quantum hall effect and multiplexed light sources of large orbital angular momenta},\ }\href {https://doi.org/10.1038/s41567-021-01165-8} {\bibfield  {journal} {\bibinfo  {journal} {Nat. Phys.}\ }\textbf {\bibinfo {volume} {17}},\ \bibinfo {pages} {700} (\bibinfo {year} {2021})}\BibitemShut {NoStop}%
\bibitem [{\citenamefont {Noh}\ \emph {et~al.}(2018)\citenamefont {Noh}, \citenamefont {Benalcazar}, \citenamefont {Huang}, \citenamefont {Collins}, \citenamefont {Chen}, \citenamefont {Hughes},\ and\ \citenamefont {Rechtsman}}]{nohTopological2018}%
  \BibitemOpen
  \bibfield  {author} {\bibinfo {author} {\bibfnamefont {J.}~\bibnamefont {Noh}}, \bibinfo {author} {\bibfnamefont {W.~A.}\ \bibnamefont {Benalcazar}}, \bibinfo {author} {\bibfnamefont {S.}~\bibnamefont {Huang}}, \bibinfo {author} {\bibfnamefont {M.~J.}\ \bibnamefont {Collins}}, \bibinfo {author} {\bibfnamefont {K.~P.}\ \bibnamefont {Chen}}, \bibinfo {author} {\bibfnamefont {T.~L.}\ \bibnamefont {Hughes}},\ and\ \bibinfo {author} {\bibfnamefont {M.~C.}\ \bibnamefont {Rechtsman}},\ }\bibfield  {title} {\bibinfo {title} {Topological protection of photonic mid-gap defect modes},\ }\href {https://doi.org/10.1038/s41566-018-0179-3} {\bibfield  {journal} {\bibinfo  {journal} {Nat. Photonics}\ }\textbf {\bibinfo {volume} {12}},\ \bibinfo {pages} {408} (\bibinfo {year} {2018})}\BibitemShut {NoStop}%
\bibitem [{\citenamefont {Deng}\ \emph {et~al.}(2022)\citenamefont {Deng}, \citenamefont {Benalcazar}, \citenamefont {Chen}, \citenamefont {Oudich}, \citenamefont {Ma},\ and\ \citenamefont {Jing}}]{deng_observation_2022}%
  \BibitemOpen
  \bibfield  {author} {\bibinfo {author} {\bibfnamefont {Y.}~\bibnamefont {Deng}}, \bibinfo {author} {\bibfnamefont {W.~A.}\ \bibnamefont {Benalcazar}}, \bibinfo {author} {\bibfnamefont {Z.-G.}\ \bibnamefont {Chen}}, \bibinfo {author} {\bibfnamefont {M.}~\bibnamefont {Oudich}}, \bibinfo {author} {\bibfnamefont {G.}~\bibnamefont {Ma}},\ and\ \bibinfo {author} {\bibfnamefont {Y.}~\bibnamefont {Jing}},\ }\bibfield  {title} {\bibinfo {title} {Observation of degenerate zero-energy topological states at disclinations in an acoustic lattice},\ }\href {https://doi.org/10.1103/PhysRevLett.128.174301} {\bibfield  {journal} {\bibinfo  {journal} {Phys. Rev. Lett.}\ }\textbf {\bibinfo {volume} {128}},\ \bibinfo {pages} {174301} (\bibinfo {year} {2022})},\ \bibinfo {note} {publisher: American Physical Society}\BibitemShut {NoStop}%
\bibitem [{\citenamefont {Ma}\ \emph {et~al.}(2019)\citenamefont {Ma}, \citenamefont {Xiao},\ and\ \citenamefont {Chan}}]{ma2019topological}%
  \BibitemOpen
  \bibfield  {author} {\bibinfo {author} {\bibfnamefont {G.}~\bibnamefont {Ma}}, \bibinfo {author} {\bibfnamefont {M.}~\bibnamefont {Xiao}},\ and\ \bibinfo {author} {\bibfnamefont {C.~T.}\ \bibnamefont {Chan}},\ }\bibfield  {title} {\bibinfo {title} {Topological phases in acoustic and mechanical systems},\ }\href@noop {} {\bibfield  {journal} {\bibinfo  {journal} {Nature Reviews Physics}\ }\textbf {\bibinfo {volume} {1}},\ \bibinfo {pages} {281} (\bibinfo {year} {2019})}\BibitemShut {NoStop}%
\bibitem [{\citenamefont {Politano}\ \emph {et~al.}(2017)\citenamefont {Politano}, \citenamefont {Viti},\ and\ \citenamefont {Vitiello}}]{politano2017optoelectronic}%
  \BibitemOpen
  \bibfield  {author} {\bibinfo {author} {\bibfnamefont {A.}~\bibnamefont {Politano}}, \bibinfo {author} {\bibfnamefont {L.}~\bibnamefont {Viti}},\ and\ \bibinfo {author} {\bibfnamefont {M.~S.}\ \bibnamefont {Vitiello}},\ }\bibfield  {title} {\bibinfo {title} {Optoelectronic devices, plasmonics, and photonics with topological insulators},\ }\href@noop {} {\bibfield  {journal} {\bibinfo  {journal} {APL Materials}\ }\textbf {\bibinfo {volume} {5}} (\bibinfo {year} {2017})}\BibitemShut {NoStop}%
\end{thebibliography}%

\end{document}